\documentclass[pdflatex,sn-mathphys,iicol]{sn-jnl}

\usepackage{amsmath}  
\usepackage{amsfonts} 
\usepackage{graphicx} 
\usepackage{color}
\usepackage{lipsum}

\newcommand {\ggtau} {g^{(2)}(\tau)}   
\newcommand {\gtau} {g^{(1)}(\tau)}
\newcommand {\tauc} {\tau_\mathrm{c}}
\newcommand {\ggomegasc} {\tilde{g}_\mathrm{sc}{}^{(2)}(\omega)}
\newcommand {\gomegasc} {\tilde{g}_\mathrm{sc}{}^{(1)}(\omega)}

\DeclareRobustCommand{\ion}[2]{\textup{#1\,\textsc{\lowercase{#2}}}}

\jyear{2021}%

\theoremstyle{thmstyleone}%
\theoremstyle{thmstyletwo}%

\theoremstyle{thmstylethree}%

\begin{document}

\title[Article Title]{Field and intensity correlations: the Siegert relation from stars to quantum emitters}


\author[1]{\fnm{Pierre Lass\`egues}}
\author[1,2]{\fnm{Mateus Antônio Fernandes Biscassi}}
\author[1]{\fnm{Martial Morisse}}
\author[2]{\fnm{Andr\'e Cidrim}}
\author[1]{\fnm{Nolan Matthews}}
\author[1]{\fnm{Guillaume Labeyrie}}
\author[3]{\fnm{Jean-Pierre Rivet}}
\author[3]{\fnm{Farrokh Vakili}}
\author[1]{\fnm{Robin Kaiser}}
\author[1]{\fnm{William Guerin}}
\author[1,2]{\fnm{Romain Bachelard}}
\author*[1]{\fnm{Mathilde Hugbart}}\email{mathilde.hugbart@inphyni.cnrs.fr}

\affil*[1]{Universit\'e C\^ote d'Azur, CNRS, INPHYNI, France}
\affil[2]{Departamento de F\'{\i}sica, Universidade Federal de S\~{a}o Carlos, Rodovia Washington Lu\'{\i}s, km 235 - SP-310, 13565-905 S\~{a}o Carlos, SP, Brazil}
\affil[3]{Universit{\'e} C{\^o}te d'Azur, Observatoire de la C{\^o}te d'Azur, CNRS, Laboratoire Lagrange, France}


\abstract{The Siegert relation relates field and intensity temporal correlations. After a historical review of the Siegert relation and the Hanbury Brown and Twiss effect, we discuss the validity of this relation in two different domains. We first show that this relation can be used in astrophysics to determine the fundamental parameters of stars, and that it is especially important for the observation with stellar emission lines. Second, we verify the validity of this relation for moving quantum scatterers illuminated by a strong driving field.}




\maketitle

\section{Introduction}\label{sec:introduction}

Light can be described using different tools, and in particular through the ones linked to its wave behaviour, such as its coherence properties. The knowledge of these properties provides information about the light source itself, such as its angular intensity profile if one measures the spatial coherence, but also on the underlying light matter interaction processes when, for example, one measures the temporal coherence of light emitted or scattered by a medium. Temporal coherence properties are often characterized through the light spectrum $S(\omega)$, which corresponds to the light intensity distribution as a function of wavelength or frequency.  More formally, for stationary processes, the spectrum is linked to the temporal field correlation function $g^{(1)}(\tau)$ through the Wiener-Khintchine theorem\,\cite{Wiener1930,Khintchine1934}:
\begin{equation}
    S(\omega) = \int g^{(1)}(\tau) e^{i\omega\tau} d\tau.
\end{equation}
The temporal field correlation function $g^{(1)}(\tau)$, also called the first-order correlation function, is defined as:
\begin{equation}\label{eq:g1}
    g^{(1)}(\tau) = \frac{\langle E^\star(t) E(t + \tau) \rangle}{\langle E^\star(t)E (t) \rangle},
\end{equation}
where $\langle . \rangle$ corresponds to the averaging over time $t$, and where the intensity of the field is given by $I(t) = E^\star(t)E (t)$. 

To have a full description of the temporal coherence properties, one has to know the correlation functions $g^{(n)}(\tau)$ at all orders $n$\,\cite{Glauber:1963b}:
\begin{eqnarray}
     &&g^{(n)}(t_1,t_2,...,t_n,t_{n+1},...,t_{2n}) = \nonumber\\
     &&\frac{\langle \Pi_{j=1}^n E^\star(t_j) \Pi_{m=n+1}^{2n} E(t_m) \rangle}{\langle I(t) \rangle^n}.\label{eq:gn}
\end{eqnarray}
However, there are cases where a full knowledge of correlation functions is not necessary. This is, for example, the case for chaotic light, such as the one emitted by a large number of independent scatterers, or more generally for light with a Gaussian distribution for the electric field. In this case, there exists a relation between the correlation functions at all orders. In particular, for spatially coherent polarized light one can relate the first orders of the correlation functions by the Siegert relation~\cite{Siegert:1943}
\begin{equation}\label{eq:Siegert_simple}
    g^{(2)}(\tau) = 1+\lvert g^{(1)}(\tau)\rvert^2,
\end{equation}
where $g^{(2)}(\tau)$ is the second-order temporal correlation function, or the temporal intensity correlation function, defined as:
\begin{equation}\label{eq:g2}
    g^{(2)}(\tau) = \frac{\langle I(t)I(t + \tau) \rangle}{\langle I(t) \rangle^2}.
\end{equation}


Physically, the $g^{(1)}(\tau)$ function quantifies the degree of mutual coherence between time-delayed electric fields: for chaotic light, it is equal to 1 at $\tau=0$ and decreases to zero at large time-delays over a characteristic time scale given by the coherence time of the field, inversely proportional to the spectrum width.
From the Siegert relation, one can deduce that the $g^{(2)}(\tau)$ function decreases from 2 to 1, and that the intensity has a coherence time which is half the one of the electric field. The excess of intensity correlation at short delay ($g^{(2)}(\tau=0)>1$) is referred to the ``bunching'' of photons or ``the Hanbury Brown and Twiss (HBT) effect'', for historical reasons described in the next section.

But first, let us give a physical picture to understand this bunching effect in the simplest case of a spatially coherent chaotic source. Consider radiation with a finite optical spectrum of linewidth $\Delta\omega$. Since the source is chaotic, there is no phase relationship between the different spectral components. This is the case, for instance, if the light is generated from many independent emitters each with different velocities. In this configuration, the spectral phase $\phi(\omega)$ can be considered as random and uniformly distributed between 0 and 2$\pi$. Let us now consider two frequency components from the optical spectrum. They induce a beat note at a frequency given by the difference between their optical frequencies. Since $\phi(\omega)$ is a random variable, all the possible beat notes coming from all possible pairs sum up with random phases. This mechanism is responsible for intensity fluctuations. Note that this is a fully classical noise, due to the wave nature of the field and to its non-monochromaticity. This noise adds up to the photon noise (shot noise), which has a quantum origin. If the linewidth of the spectrum is infinite, one would get white noise and thus an intensity correlation function equal to unity, whatever the delay $\tau$. On the other hand,
a finite linewidth means that there is no beating at a frequency much larger than $\Delta\omega$. This cut-off in the power spectrum of the noise corresponds to a finite coherence time $\tauc \sim 1/\Delta\omega$, and thus to correlated intensity fluctuations on this typical time scale. 


The Siegert relation has been used in different domains. One can cite, for example, dynamic light scattering\,\cite{Maret_1987,Pine_1988}, where the intensity correlation is analyzed to determine the size of small scatterers. In this paper, we focus on two specific domains: astronomy and light scattered by quantum particles. After a brief introduction on the history of intensity correlations, we present how the Siegert relation can be used to determine fundamental parameters of stars. This relation is in particular interesting when one observes stars with strong emission lines. We then turn to the light scattered by quantum particles, namely cold atoms. We show that when those atoms are illuminated by a strong driving field, the incoherent scattered light still verifies the Siegert relation. Whereas this relation is usually derived in the classical domain \cite{Loudon:book,Chu2012}, we give in Section \ref{sec:derivation} the detailed derivation for quantum emitters.

\section{A brief history of intensity correlations}\label{sec:history}

The history of the Siegert relation is intimately linked to the controversy on equal-time intensity autocorrelations, also known as the Hanbury~Brown and Twiss effect. The story starts during World War II, when radar technology drove a lot of research in the field of radio waves with, later, much repercussion on radio astronomy and optical sciences. The relation between electric field correlations and intensity correlations has been proposed in that context by A. J. F. Siegert~\cite{Siegert:1943} in a technical report. It was later named ``Siegert relation'', mainly in the field of mesoscopic physics\,\cite{Teich_1988,akkermans_montambaux_2007}.

\begin{figure}[t]
	\centering
	\includegraphics[width=\columnwidth]{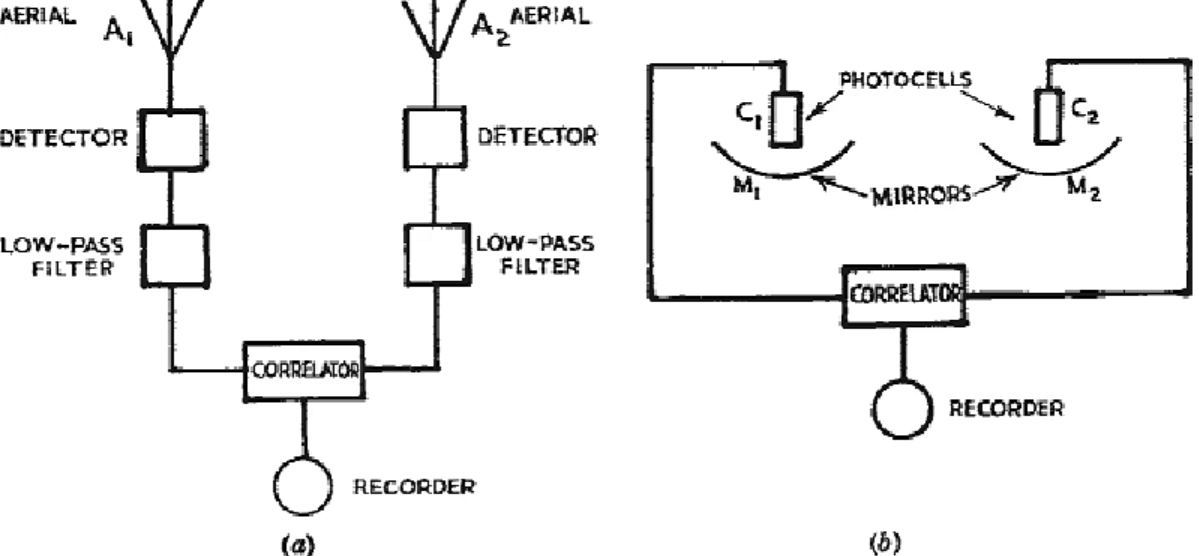}
	\caption{Simplified outline of an intensity interferometer for radio waves (a) and optical frequencies (b) taken from Ref.\,\cite{HBT:1956a}}
	\label{fig:HBT_1956}
\end{figure}

The next important step has been achieved by Hanbury~Brown and Twiss in the field of radio astronomy. In 1952, they proposed and demonstrated a novel type of radio interferometer. 
The intensities of radio waves collected by two telescopes were recorded and correlated (see Figure.~\ref{fig:HBT_1956}), first for the Sun and then subsequently for two radio sources in Cygnus and Cassiopeia, without measuring the electromagnetic amplitude and phase information. The angular sizes of these sources were determined by measuring the spatial intensity correlations for several different configurations of telescope separations~\cite{HBT:1952}. In their 1954 paper\,\cite{HBT:1954} they wrote: `\textit{It is further shown that the correlator output, when suitably normalized, is equal to the square of the correlation coefficient measured by the Michelson interferometer}'. This statement links the intensity correlation function with the field correlation function, establishing again the Siegert relation. Indeed the original technical report by Siegert had remained largely unnoticed and the relation was independently rediscovered at that time.

It was then natural, for radio-astronomers, to extend this new concept to visible light. However, they faced a strong opposition from several physicists who preferred to think about light in terms of photons~\cite{HB:Boffin}. Indeed, a temporal (or spatial) intensity correlation measurement relies on the detection of at least two photons. The classical description given in the previous section only relies on interference. What was puzzling at that time is that physicists were convinced, following Dirac, that ``\textit{interference between two different photons can never occur}'' \cite{Dirac:1930}.

HBT successfully tested their idea with a laboratory demonstration~\cite{HBT:1956a}, and a few months later on the light from a star~\cite{HBT:1956b}. Their results were first disputed as other groups failed to reproduce the lab experiment, and as it was claimed that such results, if true, would call for a major revision of quantum mechanics~\cite{Brannen:1956}. Nevertheless, it was later shown that
the other experiments were simply not sensitive enough~\cite{HBT:1956c}. These first experiments were performed with continuous wave detection, in which case the notion of photons is not needed and the classical explanation is perfectly appropriate. In the photon-counting regime, one can still assume that the quantization only occurs at the detection of an underlying continuous quantity, in which case the instantaneous value of the intensity $I(t)$ gives the probability of detecting photons, and the classical picture is still valid.

However, if one insists on describing light in terms of photons, another physical description is needed. A first argument was given by Purcell in 1956\,\cite{Purcell:1956}: the bunching of photons is a consequence of the Bose-Einstein statistics to which they obey. This interpretation was further developed by HBT~\cite{HBT:1957a} and Kahn~\cite{Kahn:1958}, and verified in an experiment done, again by HBT, in the photon-counting regime~\cite{HBT:1957}. Finally, another interpretation in terms of two-photon interference was given by Fano a few years later~\cite{Fano:1961}.

The HBT experiment and its understanding in the framework of quantum mechanics can be considered as the birth of modern quantum optics (before the laser was invented!). In particular, and as acknowledged by Glauber in his Nobel prize speech~\cite{Glauber:2006}, it triggered the development of the quantum theory of optical coherence~\cite{Glauber:1963a,Glauber:1963b,Glauber:1963c}, based in particular on correlation functions. In this context, the Siegert relation, which provides a relation between the first and second-order field correlations, is a particularly important tool to probe the quantum nature of light. Intensity correlations experiments have been used as a signature to distinguish laser light from classical fields\,\cite{Arecchi:1966}. This experiment illustrates that higher order photon statistics are a fundamental tool to identify non classical states of light, which could not be identified as such by field-field correlation functions or the optical spectrum. The use of intensity correlation functions have also allowed the pioneering experiments illustrating the violation of Bell inequalities and effects on single photon sources in the early 80s\,\cite{Aspect_1981, Aspect_1982_2, Aspect_1982, Grangier_1985} which opened the path towards the area of modern quantum technologies.

\section{The Siegert relation in astronomy}\label{sec:Stars}


The use of intensity correlations, also called intensity interferometry in astronomy, was pioneered by HBT, who measured the angular diameters of 32 stars and their fundamental
characteristics from the spatial intensity correlation function\,\cite{HBT1974}. After this series of impressive observations using the Narrabri Stellar Intensity Interferometer in Australia in the 1960s-1970s, this technique has been abandoned, mainly due to its poor signal-to-noise ratio compared to emerging techniques pioneered by A. Labeyrie in amplitude interferometry\,\cite{Labeyrie:1975}. However, thanks to progress in photonics components, efficient single photon counting detectors, fast electronics and digital correlators, there is currently a strong effort from different groups towards the revival of intensity interferometry with modern photonic technologies. 
One strategy is to utilize large diameter ($> 10\,$m) imaging air Cherenkov telescopes for intensity interferometry~\cite{LeBohec:2006} and has recently resulted in successful on-sky measurements~\cite{Acciari:2020,Abeysekara:2020}. Our team is following an alternative approach by using traditional astronomical optical telescopes with photon-counting avalanche photodiodes (APDs) feeding a fast time-tagger, which then computes the temporal correlations in real time~\citep{Guerin:2018, Rivet:2018, Lai:2018, Matthews:2022}.

Note that astronomers are mainly interested in the \emph{spatial} intensity correlation function. While in the temporal domain, the bunching is related to the spectrum width of the source via the Siegert relation and the Wiener-Khintchine theorem, in the spatial domain, the bunching (which can be seen as the typical size of a speckle grain) is similarly related to the angular width of the source via the van Cittert-Zernike theorem \cite{Goodman:book}. In the following, however, we will focus on temporal intensity correlation measurements and show how this can be used to determine fundamental parameters of stars.


\subsection{Experimental setup}

\begin{figure}[b]
	\centering
	\includegraphics[width=\columnwidth]{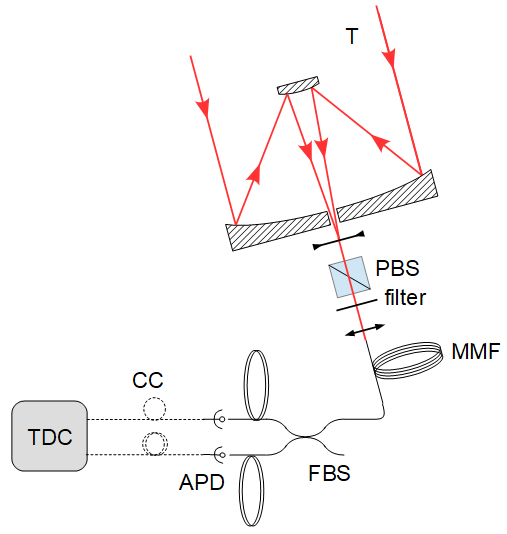}
	\caption{Experimental setup to measure the temporal intensity correlation for light coming from stars. See text for details. T: telescope, PBS: Polarizing Beam Splitter, MMF: multimode fiber, FBS: fibered beamsplitter, APD: avalanche photodiodes, CC: $50 \Omega$ coaxial cables, TDC: time-to-digital convertor.
     }
	\label{fig:Setup_Telescope}
\end{figure}

The light coming from stars is generally assumed to be chaotic, so that it satisfies the Siegert relation. In order to test this assumption, one needs to measure both $g^{(2)}(\tau)$ and $g^{(1)}(\tau)$.  $S(\omega)$ can also be measured instead of $g^{(1)}(\tau)$, since they are related through the Wiener-Khintchine theorem. In our specific case, the $g^{(2)}(\tau)$ function is measured thanks to the simplified setup depicted in Fig.\,\ref{fig:Setup_Telescope} and described in details in Ref.\,\cite{Almeida2022}. Very briefly, the light is first collected by a 1\,m telescope at the C2PU facility on the Plateau de Calern site of Observatoire de la C\^ote d’Azur (OCA). It is then injected in a coupling assembly to perform spectral filtering with a narrow-band interference filter, polarization filtering with a polarizing beam splitter (PBS), and then injection in a 100\,$\mu$m core multimode fiber (MMF). Since the stars we are looking at are not resolved by the telescope, we still inject only one spatial mode into the fiber. Finally, the output of the fiber is connected to a fibered beamsplitter (FBS) whose outputs illuminate two single-photon avalanche photodiodes (APDs). The photons detected by the APDs are time-tagged by a time-to-digital convertor (TDC) that also computes the temporal intensity correlation.

\subsection{Test in the laboratory}

\begin{figure}[t]
	\centering
	\includegraphics[width=\columnwidth]{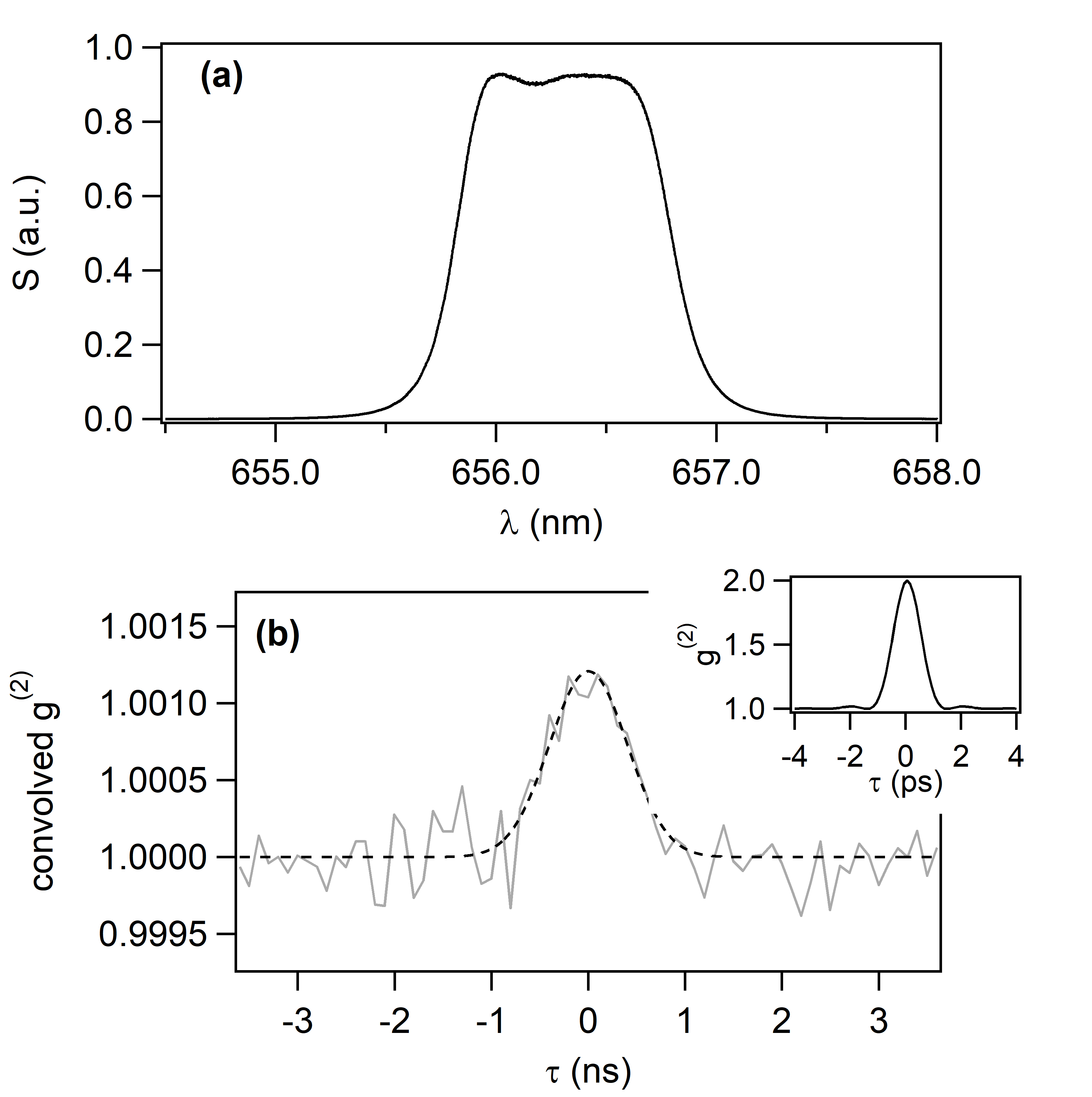}
	\caption{(a) Spectrum of the light coming from an halogen lamp injected in a monomode fiber (artificial star) and transmitted by a narrow filter, showing a spectral bandwidth of about 1\,nm. (b) Grey line: temporal $g^{(2)}$ function measured on the artificial star. Dashed line: Theoretical $g^{(2)}(\tau)$ function, calculated from the spectrum assuming the Siegert relation, convolved by the electronic time response. Inset: theoretical $g^{(2)}(\tau)$ function calculated from the transmitted spectrum, assuming the Siegert relation and an infinite electronic response bandwidth.}
	\label{fig:Siegert_Astro_Lab}
\end{figure}

The Siegert relation can be first demonstrated in the lab using an artificial unresolved star. The source is generated by injecting light from a halogen lamp, with a broad continuous spectrum, into a single mode fiber. The spectrum of the light on which $g^{(2)}(\tau)$ is computed thus corresponds to the spectrum of the light after the filter. This transmitted spectrum is plotted in Fig.\,\ref{fig:Siegert_Astro_Lab}a. The bandwidth of the filter is $\Delta \lambda = 1$\,nm with a central frequency $\lambda_0=656.3$\,nm.

From the spectrum transmitted by the filter, we can calculate its Fourier transform $g^{(1)}(\tau)$ and then $g^{(2)}(\tau)$ using the Siegert relation. The result is plotted in the inset of Fig.\,\ref{fig:Siegert_Astro_Lab}b. The coherence time $\tau_\mathrm{c}$ can be defined as the area of the bunching peak\,\cite{Mandel1965}. For technical reasons that will be explained in the following, we focus on the area, equal to 1.19\,ps for this lab test. This coherence time is much lower than the electronic time resolution, mainly limited by the APD jitter, which is of the order of 500\,ps, but which depends on different parameters such as the count rate and the beam size on the APD. This broader time response reduces the height of the measured bunching peak by approximately the ratio of the coherence time to the resolving time. On the other hand, the bunching width increases and its area remains constant.

The temporal jitter function can be assumed to be Gaussian at first order. Its full-width at half-maximum (FWHM) is estimated to be equal to $\sqrt2 \times 670$\,ps for this specific measurement, the $\sqrt2$ factor being due to the fact that we have two APDs to perform the measurement. The theoretical $g^{(2)}$ function now corresponds to the theoretical bunching peak with infinite electronic response bandwidth, shown in the inset of Fig.\,\ref{fig:Siegert_Astro_Lab}b, convolved by the Gaussian time response. This convolved $g^{(2)}$ function is shown in Fig.\,\ref{fig:Siegert_Astro_Lab}b with the dashed line. The area is still equal to 1.19\,ps.

Finally, we plot the measured temporal $g^{(2)}$ function. This corresponds to the grey line in Fig.\,\ref{fig:Siegert_Astro_Lab}b. One can see that it is perfectly superimposed to the convolved theoretical $g^{(2)}$ function, as expected from the Siegert relation. The area measured with a Gaussian fit applied on the measured data gives $1.18\pm0.08$, in agreement with the expected 1.19\,ps. With a much longer integration time and a precise experimental characterization of the filtered spectrum, we have recently obtained an excellent agreement to the 1\% level\,\cite{Matthews:2022}.

\subsection{Measurements on stars}

The next step is to check the Siegert relation for on-sky measurements. The first measurements were performed in the continuum of $\alpha$\,Boo (Arcturus), $\alpha$\, CMi (Procyon) and $\beta$ Gem (Pollux), with a filter centered at 780\,nm\,\cite{Guerin:2017}. Since we are in the continuum, the spectra of the stars can be considered as constant over the transmission of the filter, and the bunching peak is thus simply limited by the bandwidth filter. In other words, the measured coherence time is limited by the filter and does not give any information on the star itself.

More interestingly, we can also focus on spectral emission lines. O, B and A supergiant stars have massive winds witnessed by those strong emission lines (e.g. hydrogen Balmer series, helium or carbon lines among others in the visible), and they mainly originate in a circumstellar envelope that extends from a few to tens and even hundreds of stellar radii. A well-known example in the northern hemisphere is P\,Cygni, which has a strong H$\alpha$ emission line at 656.3\,nm. To check the validity of the Siegert relation, one needs, as done in the previous section with the artificial star, to compute the spectrum transmitted by the filter thanks to the filter transmission and the measured star spectrum, and to measure the $g^{(2)}$ function of the light collected by one telescope. Since the star is not resolved by the telescope, we can consider that only one spatial mode is collected.

\begin{figure}[t]
	\centering
	\includegraphics[width=\columnwidth]{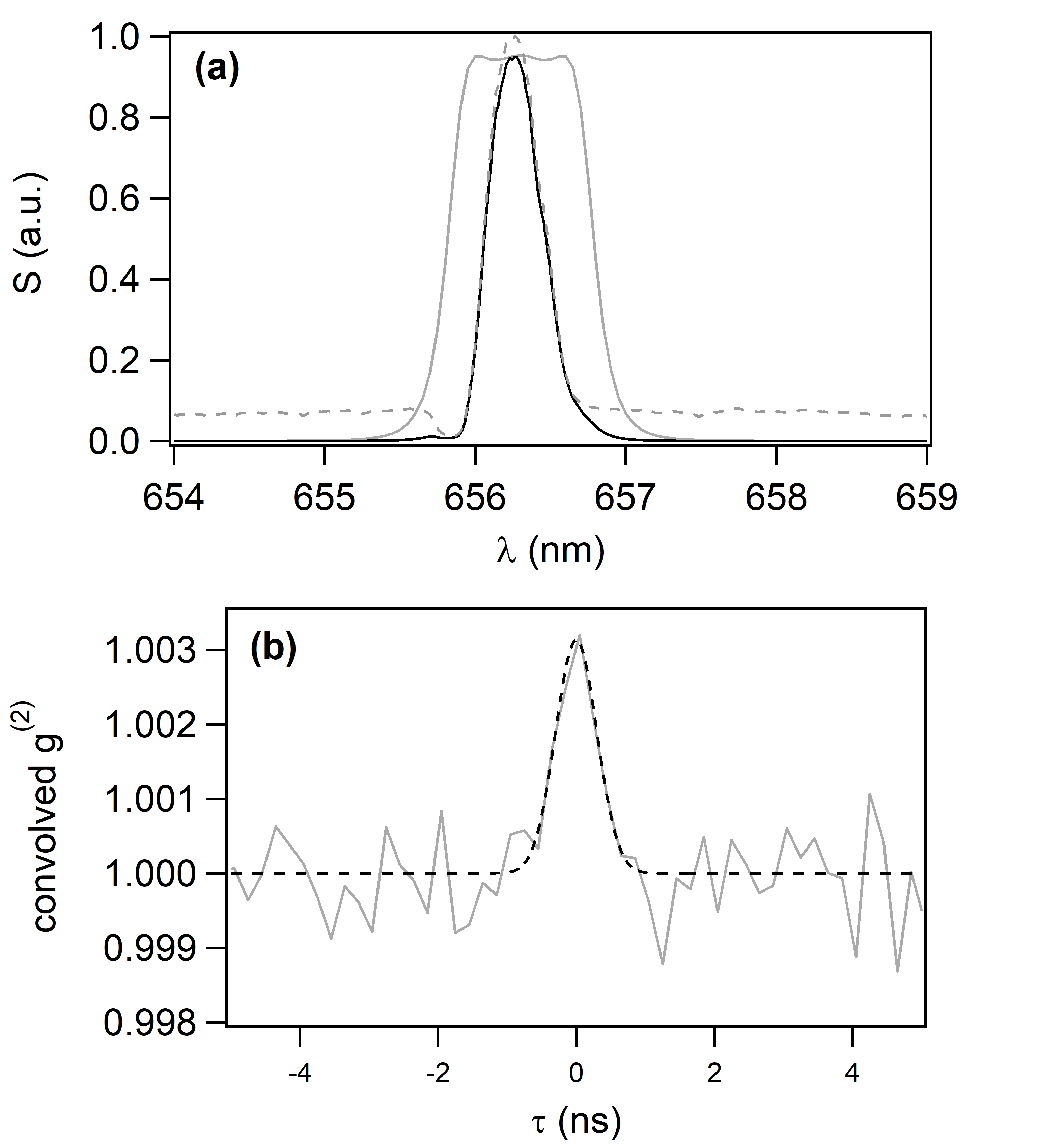}
	\caption{(a) Grey line: H$\alpha$ filter transmission. Dashed grey line: spectrum emitted by P\,Cygni as reported in the AAVSO database\,\cite{AAVSO} on August 8th 2020. Black line: spectrum of P\,Cygni transmitted by the filter. (b) Grey line: temporal $g^{(2)}$ function measured on P\,Cygni between 3 August 2020 and 9 August 2020. Dashed line: Theoretical $g^{(2)}(\tau)$ function, calculated from the spectrum assuming the Siegert relation and convolved by the electronic time response.}
	\label{fig:Siegert_Pcygni}
\end{figure}

These lines exhibit time variability at different scales, from days to years and beyond. It is thus important to measure the stellar spectrum in the same period as the $g^{(2)}$ measurement. Fig.\,\ref{fig:Siegert_Pcygni}(a) presents the spectrum emitted by P\,Cygni observed on August 8$^\mathrm{th}$ 2020\,\cite{AAVSO}. One can see that the linewidth, of the order of 0.5\,nm, is smaller than the 1\,nm width of the filter, with an emission line 15 times higher than the continuum. As already done in the previous section, we calculate the convolved $g^{(2)}$ from this spectrum, with a FWHM of the time response equal to 500\,ps. This result is superimposed in Fig.\,\ref{fig:Siegert_Pcygni}(b) to the $g^{(2)}$ measurements performed on-sky between 3 August 2020 and 9 August 2020\,\cite{Almeida2022}. Again,  a good overlap is observed, illustrating the validity of the Siegert relation. The area of the theoretical $g^{(2)}$ function is equal to 2.35\,ps, also in agreement with the area extracted from a Gaussian fit of the measurement, equal to $2.3\pm\,0.3$\,ps.

We thus show here that the Siegert relation is valid for the light emitted by stars, in the continuum but also for specific cases such as strong emission lines. The measurement of simultaneous spectra and spatial intensity correlation functions $g^{(2)}(r)$ with two telescopes, the correlation at zero baseline being given by $g^{(2)}(\tau)$ measured with one telescope, is a powerful tool to estimate the angular diameter of the star, as pioneered by HBT. In association with spectral measurement and radiative transfer modelling, the distance of the star can be inferred, as shown in Refs.\,\cite{Rivet:2020,Almeida2022}.

\subsection{Towards quantum astronomy}

The fact that starlight fulfills the Siegert relation is obviously not a surprise. Nevertheless, one may find, in the future, some natural sources of light in space for  which this would not be necessarily true. Indeed, it has been speculated that natural astrophysical ``lasers'' (emission lines amplified by stimulated emission) could exist \cite{Johansson:2007}.

In the microwave domain, space masers are very common \cite{Reid:1981, Elitzur:1982}, but as the wavelength decreases, obtaining a population inversion gets harder.
Amplification by stimulated emission was indeed observed in the planetary atmospheres of Mars and Venus in the infrared (IR) ($\lambda \sim 10\,\mu$m) \cite{Johnson:1976,Mumma:1981} and much later in stellar atmospheres in the far IR \cite{Strelnitski:1996}. More recently, the discovery of astrophysical lasers in the near IR was claimed by Johansson and Letokhov, based on \ion{Fe}{II} \cite{Johansson:2004} and \ion{O}{I} \cite{Johansson:2005}, but this interpretation is disputed \cite{Messenger:2010}. The most promising emission line seems to be the \ion{Fe}{II} line at 1.68$\mu$m, which should be present in the circumstellar envelope of some spectral types of stars \cite{Messenger:2010}.

One could even imagine that scattering-induced feedback could enhanced the amplification, even reaching the oscillatory regime \cite{Lavrinovich:1975}, a phenomenon that would be called a random laser today \cite{Wiersma:2008,Baudouin:2013}. Measuring the $g^{(2)}(\tau)$ function of such exotic emission lines would thus open the way to the study of coherence and quantum effects in space and intensity correlation experiments performed in laboratory random lasing systems indicate this to be a convenient indicator\,\cite{Cao_2001, Raposo_2022}.


\section{The Siegert relation for quantum emitters}\label{sec:Quantum}

We now turn to the Siegert relation for light scattered by quantum scatterers. As will be discussed later, our scattering medium is made of cold Rb atoms, illuminated by a laser probe beam. Varying the intensity of this probe allows us to tune the saturation parameter $s$, and thus the ratio between elastic and inelastic scattering. For elastic scattering, the different moving scatterers each independently shift the frequency of the incident light by Doppler effect, so the scattered field has the properties of the incident one, yet broadened\,\cite{Eloy_2018}. In this case, we can consider the scatterers as classical ones and they generate what is often called pseudothermal light\,\cite{Kuusela_2017}. To validate the Siegert relation, one needs a large number of independent scatterers and a process to randomize the phase between the fields emitted by the different scatterers. For elastic scattering, the phase is randomized thanks to nonzero temperature. The validity of the Siegert relation in this configuration is presented in Ref.\,\cite{Ferreira_2020}.

When the saturation parameter is much larger than one, the scattered field is dominated by the inelastic one, which corresponds to the well-known Mollow triplet for the spectrum\,\cite{Mollow_1969}. This effect is purely quantum, the fluorescence spectrum corresponding to the one of a two-level system driven by a strong and resonant incident field. The Mollow triplet is an effect for which strong correlations exist between
the two sidebands and the carrier\,\cite{Schrama_1992}, as first observed by A. Aspect et al. in 1980\,\cite{Aspect_1980}. The existence of correlations in the photon radiation also leads to antibunching, usually observed for a single two-level emitter, but which can be observed in large systems with a phase-matching experimental configuration\,\cite{Grangier_1986}. Even if not obvious, the Siegert relation is still valid in our setup, as will be demonstrated both with its derivation and experimental results in the following.

\subsection{Derivation in the quantum formalism}\label{sec:derivation}

For the derivation of the Siegert relation, we consider a large number of uncorrelated scatterers. A quantum treatment of the correlation functions $g^{(1)}$ and $g^{(2)}$ requires the use of electric field operators $\hat{E}$, carefully accounting for their ordering, also known as ``normal ordering''~\cite{Glauber:1963b}. This also requires a quantum treatment of the emitters, here modelled as two-level atoms. Furthermore, the derivation of the Siegert relation requires that the mean value, which corresponds to the expected value in the quantum case, of the single-atom electric field vanishes. 
For two-level atoms, the electric field operator $\hat{E}_j^{+}$ of atom $j$ is proportional to the atomic lowering operator $\hat\sigma_j$, which is characterized by a fluctuation operator $\delta\hat{\sigma}_j=\hat\sigma_j-\langle\hat\sigma_j\rangle_{t\rightarrow\infty}$, see Ref.~\cite{steck2007quantum}. By definition, the expectation value of the fluctuations is equal to zero (yet not its variance).
The scattered electric field is given by the sum of the contribution of each scatterer $\delta\hat\sigma_j$, so the (non-normalized) correlation function for the inelastically scattered intensity can be written as:
\begin{equation}
    G^{(2)}(\tau) = \sum_{ijkl}\langle \delta\hat\sigma_i^\dagger(t)\delta\hat\sigma_j^\dagger(t\!+\!\tau)\delta\hat\sigma_k(t\!+\!\tau)\delta\hat\sigma_l(t)\rangle.
\end{equation}
We consider the scatterers to be independent from each other, so the expectation value may be split for different indexes ($\langle\hat{A}_j\hat{B}_m\rangle=\langle\hat{A}_j\rangle\langle\hat{B}_m\rangle$ for $j\neq m$). Since the expectation value of fluctuation operators is zero, most terms vanish. When one expands this equation, only the terms in which the operator acts on its complex conjugate remain:
\begin{equation}
\begin{aligned}
    G^{(2)}(\tau) =\sum\limits_{i}&\langle\delta\hat\sigma^\dagger_i(t)\delta\hat\sigma^\dagger_i(t\!+\!\tau)\delta\hat\sigma_i(t\!+\!\tau)\delta\hat\sigma_i(t)\rangle \\
     +\sum\limits_{{i,j\neq i}}
    \Big(&\langle\delta\hat\sigma^\dagger_i(t)\delta\hat\sigma^\dagger_j(t\!+\!\tau)\delta\hat\sigma_j(t\!+\!\tau)\delta\hat\sigma_i(t)\rangle \\
    +&\langle\delta\hat\sigma^\dagger_i(t)\delta\hat\sigma^\dagger_j(t\!+\!\tau)\delta\hat\sigma_i(t\!+\!\tau)\delta\hat\sigma_j(t)\rangle\Big).    
\end{aligned}
\end{equation}
Since there are only terms of the form $\hat{A}^\dagger\hat{A}$, the specific phase of the driving field on each atom is unimportant, and all scatterers can be treated equally (we neglect shadow and multiple scattering effects since the saturation parameter is very large and the optical thickness is low). Thus we can write:
\begin{equation}    
\begin{aligned}
 G^{(2)}(\tau)= N\langle\delta\hat\sigma^\dagger(t)&\delta\hat\sigma^\dagger(t+\tau)\delta\hat\sigma(t+\tau)\delta\hat\sigma(t)\rangle\\
+N(N-1&)\Big(\lvert\langle\delta\hat\sigma^\dagger(t)\delta\hat\sigma(t+\tau)\rangle\rvert^2\\&+\langle\delta\hat\sigma^\dagger(t)\delta\hat\sigma(t)\rangle^2\Big).
\end{aligned}\label{Eq:G2_unorm}
\end{equation}
For a large number of scatterers, the term scaling as $N^2$ dominates, and the other terms become negligible.
Normalizing Eq.~\eqref{Eq:G2_unorm} by the squared intensity, see Eq.~\eqref{eq:gn} where the first ($n=1$) and second ($n=2$) order coherence are computed using the operator electric field, we obtain the Siegert relation for the inelastically scattered light:
\begin{equation}
\begin{aligned}
    g^{(2)}(\tau) &= 1+\frac{\lvert\langle\delta\hat\sigma^\dagger(t)\delta\hat\sigma(t+\tau)\rangle\rvert^2}{\langle\delta\hat\sigma^\dagger(t)\delta\hat\sigma(t)\rangle^2}\\
    &= 1 +\lvert g^{(1)}(\tau)\rvert^2.
\end{aligned}    
\end{equation}
We note that the derivation is very similar to the classical case, although the operator nature of the electric field requires a careful ordering of the operators involved. Finally, the zero average value of the electric field scattered by single atoms here stems from the nature of the inelastic scattering, rather than macroscopic features such as collisions or temperature~\cite{Loudon:book}.

\subsection{Experimental setup}

In our experiment, the scattering medium corresponds to a cold atomic cloud. This medium is produced by loading a magneto-optical trap (MOT) from a vapor of $^{85}$Rb, as described in Refs.\,\cite{Kashanian_2016,Eloy_2018, Ortiz_2019,Ferreira_2020}, and then released for a given time of flight. The number of atoms $N$ is typically of the order of $10^8-10^9$. The number of scatterers is thus large, fulfilling one of the conditions needed to mimic chaotic light and thus to validate the Siegert relation. The temperature $T$ is of the order of 100\,$\mu$K. The atomic cloud is finally characterized by its on-resonance optical thickness $b_0$, measured by recording the transmission of a small probe beam going through the cloud as a function of the detuning on the $\lvert 3\rangle \rightarrow \lvert 4'\rangle$ D$_2$ hyperfine transition. In this paper, this parameter is lower than one to avoid any collective effects.

\begin{figure}[t]
	\centering
	\includegraphics[width=\columnwidth]{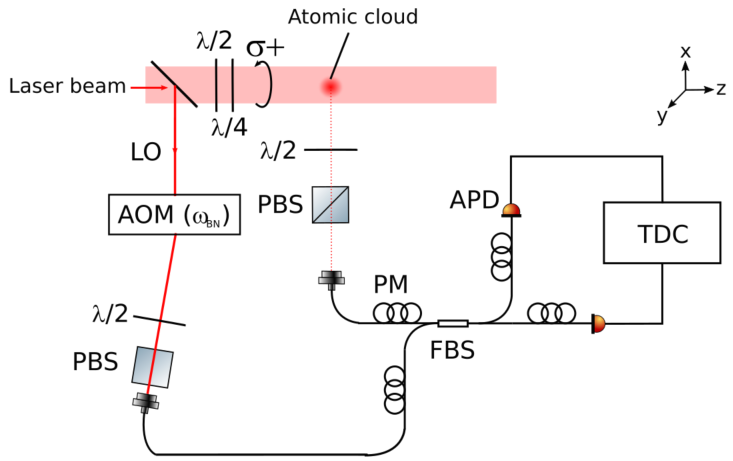}
	\caption{Experimental setup to simultaneously measure $g^{(1)}(\tau)$ and $g^{(2)}(\tau)$ of the light scattered by quantum scatterers. A cold atomic cloud is illuminated by a laser beam circularly polarized thanks to $\lambda/2$ and $\lambda/4$ plates. The scattered light is collected by a polarization-maintaining (PM) single-mode fiber after polarisation selection with a $\lambda/2$ and a polarizing beam splitter (PBS). The light is split with a fibered beam splitter (FBS) and its two outputs illuminate two avalanche photodiodes (APDs). Each photon arrival is time-tagged by a time-to-digital converter (TDC) and the correlations are computed by a computer.  Finally, a local oscillator (LO), derived from the same laser that illuminates the scattering medium and frequency-shifted by an acousto-optical modulator (AOM), is injected in the second input of the FBS.}
	\label{fig:Setup}
\end{figure}

The experimental setup is depicted in Fig.\,\ref{fig:Setup}. The cloud is illuminated by a flat-top intensity profile provided by a fibered laser (EYLSA from Quantel laser) injected in a beam shaper from Asphericon (TopShape model). Its frequency is locked on the $\lvert 3\rangle \rightarrow \lvert 4'\rangle$ hyperfine transition of the $^{85}$Rb D$_2$ line. We use waveplates to get a circularly polarized light. The beam diameter, equal to 14.7\,mm, is adjusted to be larger than the cloud radius in order to get a uniform intensity on the atoms, and thus a constant Rabi frequency. Its intensity can be changed to tune the saturation parameter between less than 0.01, for which scattering is mainly elastic, to more than 60 with mainly inelastic scattering. The duration of the probe is adjusted to limit the number of photons scattered per atoms to a few hundreds and thus limit heating effect.

The scattered light is collected at typically 90$^\circ$ from the probe beam, by a polarization-maintaining (PM) single-mode fiber. This ensures the selection of a single spatial mode. The polarization is also selected before the fiber with a $\lambda/2$ plate and a polarization beam splitter, to maximize the amount of collected photons as well as to adjust the incident polarization parallel to the PM fiber axis. To measure $\gtau$, we superimpose the collected scattered light to a local oscillator (LO) with a FBS. This local oscillator is derived from the laser which delivers the probe beam, frequency shifted by $\omega_\mathrm{BN}$ with an acousto-optical modulator. Its polarization is also adjusted before the entrance of the fiber to correspond to the PM fiber axis. The polarizations of the collected scattered light and the LO are thus parallel. The output of the FBS illuminates two APDs which are connected to a TDC. This last device allows to time-tag the arrival of each photon from which the intensity correlation function is finally calculated.

The intensity $I_\mathrm{BN}$ of the beat note between the LO and the collected scattered light is used to compute the intensity correlation function $g_\mathrm{BN}{}^{(2)}(\tau)$ in our setup:
\begin{eqnarray}
    &&g_\mathrm{BN}{}^{(2)}(\tau) = \frac{\langle I_\mathrm{BN}(t)I_\mathrm{BN}(t + \tau) \rangle}{\langle I_\mathrm{BN}(t) \rangle^2},\\
    &&\simeq 1 + 2\frac{\langle I_\mathrm{sc} \rangle \langle I_\mathrm{LO} \rangle}{\left(\langle I_\mathrm{sc} \rangle \langle I_\mathrm{LO} \rangle \right)^2}g_\mathrm{sc}{}^{(1)}(\tau) \cos(\omega_\mathrm{BN} \tau +\pi) \nonumber\\
    &&+ \frac{\langle I_\mathrm{sc} \rangle^2}{\left(\langle I_\mathrm{sc} \rangle + \langle I_\mathrm{LO} \rangle \right)^2}\left(g_\mathrm{sc}{}^{(2)}(\tau)-1 \right), \label{eq:g2_BN}
\end{eqnarray}
where $I_\mathrm{BN}$, $I_\mathrm{sc}$ and $I_\mathrm{LO}$ correspond to the intensity of the beat note, the collected scattered light and the LO, respectively. The $g_\mathrm{sc}{}^{(1)}(\tau)$ function corresponds to the temporal electric field correlation function of the scattered light, and $g_\mathrm{sc} {}^{(2)}(\tau)$ to its temporal intensity correlation function. The detailed derivation of this equation can be found in Ref.\,\cite{Ferreira_2020}. As can be seen in Eq.\,(\ref{eq:g2_BN}), this setup allows measuring the electric field and the intensity correlation functions of the light under study at the same time. This is particularly suited to check the validity of the Siegert relation, avoiding in particular any drift effect as when the $g^{(1)}$ and $g^{(2)}$ functions are measured separately.

\subsection{Results}

\begin{figure}[b]
	\centering
	\includegraphics[width=\columnwidth]{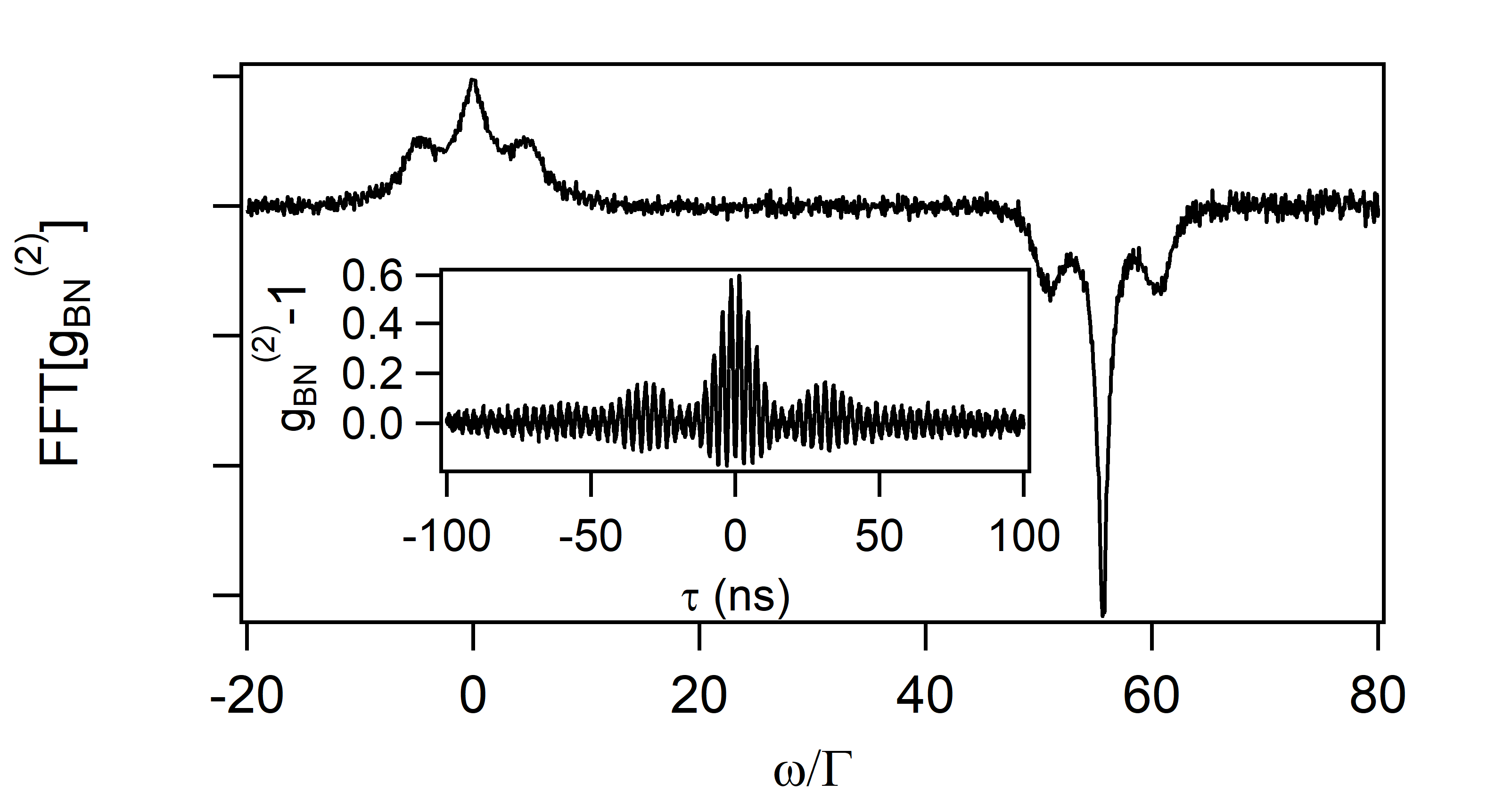}
	\caption{Fourier transform of the temporal intensity correlation function of the scattered light beating with the LO. The saturation parameter of the probe beam was set to 60. The curve close to the DC value corresponds to the Fourier transform of $g_\mathrm{sc}{}^{(2)}(\tau)$, while the frequency shifted curve corresponds to the Fourier transform of $g_\mathrm{sc}{}^{(1)}(\tau)$. Inset: temporal intensity correlation function of the scattered light beating with the LO.}
	\label{fig:g2_BN}
\end{figure}

An example of the temportal $g_\mathrm{BN}{}^{(2)}$ function obtained for $s=60$ is plotted in the inset of Fig.\,\ref{fig:g2_BN}. One observes a low frequency oscillation corresponding to the beating between the carrier and the sidebands of the Mollow triplet, as well as a fast oscillation corresponding to the beating between the scattered light and the LO. To extract the contribution of $g_\mathrm{sc}{}^{(1)}(\tau)$ and $g_\mathrm{sc}{}^{(2)}(\tau)$, we take the Fourier transform of $g_\mathrm{BN}{}^{(2)}(\tau)$. Indeed, while the Fourier transform of $g_\mathrm{sc}{}^{(2)}(\tau)$, i.e. $\ggomegasc$, is centered around the DC value, the Fourier transform of $g_\mathrm{sc}{}^{(1)}(\tau)$, $\gomegasc$, is frequency shifted by $\omega_\mathrm{BN}$. Fig.\,\ref{fig:g2_BN} presents the corresponding signal in the Fourier space. As soon as $\ggomegasc$ does not overlap with $\gomegasc$, the two quantities can be extracted separately.

The Siegert relation given by Eq.\,(\ref{eq:Siegert_simple}) in the temporal domain can be written in the Fourier space:
\begin{equation}
    \tilde{g}^{(2)}(\omega) = \delta(0)+\tilde g^{(1)}(\omega) * \tilde g^{(1)\star}(\omega).
\end{equation}
To check its validity on our experimental setup, we use the following procedure. We first correct the Fourier transform to take into account the limited temporal response of the APDs, which can be approximated by a first-order low pass filter with a bandwidth limited by the inverse of the APDs jitter. We then  extract $\ggomegasc$ and $\gomegasc$. We shift $\gomegasc$ back to zero frequency and we calculate its self-convolution. We finally renormalize this last quantity in order to match the height of $\ggomegasc$ to the one of $\gomegasc * \tilde g_\mathrm{sc}{}^{(1)\star}(\omega)$\footnote{The Fourier transforms are all real so $\gomegasc * \tilde g_\mathrm{sc}{}^{(1)\star}(\omega) = \gomegasc * \tilde g_\mathrm{sc}{}^{(1)}(\omega)$. The quantity $\gomegasc$ is negative, as can be seen in Fig.\,\ref{fig:g2_BN}, due to the $\pi$ shift in the cosine in Eq.\,(\ref{eq:g2_BN}) coming from the $\pi$ phase between the two outputs of the FBS. This negative sign is removed during the normalisation.}. Note that this normalisation depends on the intensity imbalance between $\langle I_\mathrm{sc} \rangle$ and $\langle I_\mathrm{LO} \rangle$, as can be seen in Eq.\,(\ref{eq:g2_BN}).

\begin{figure}[t]
	\centering
	\includegraphics[width=\columnwidth]{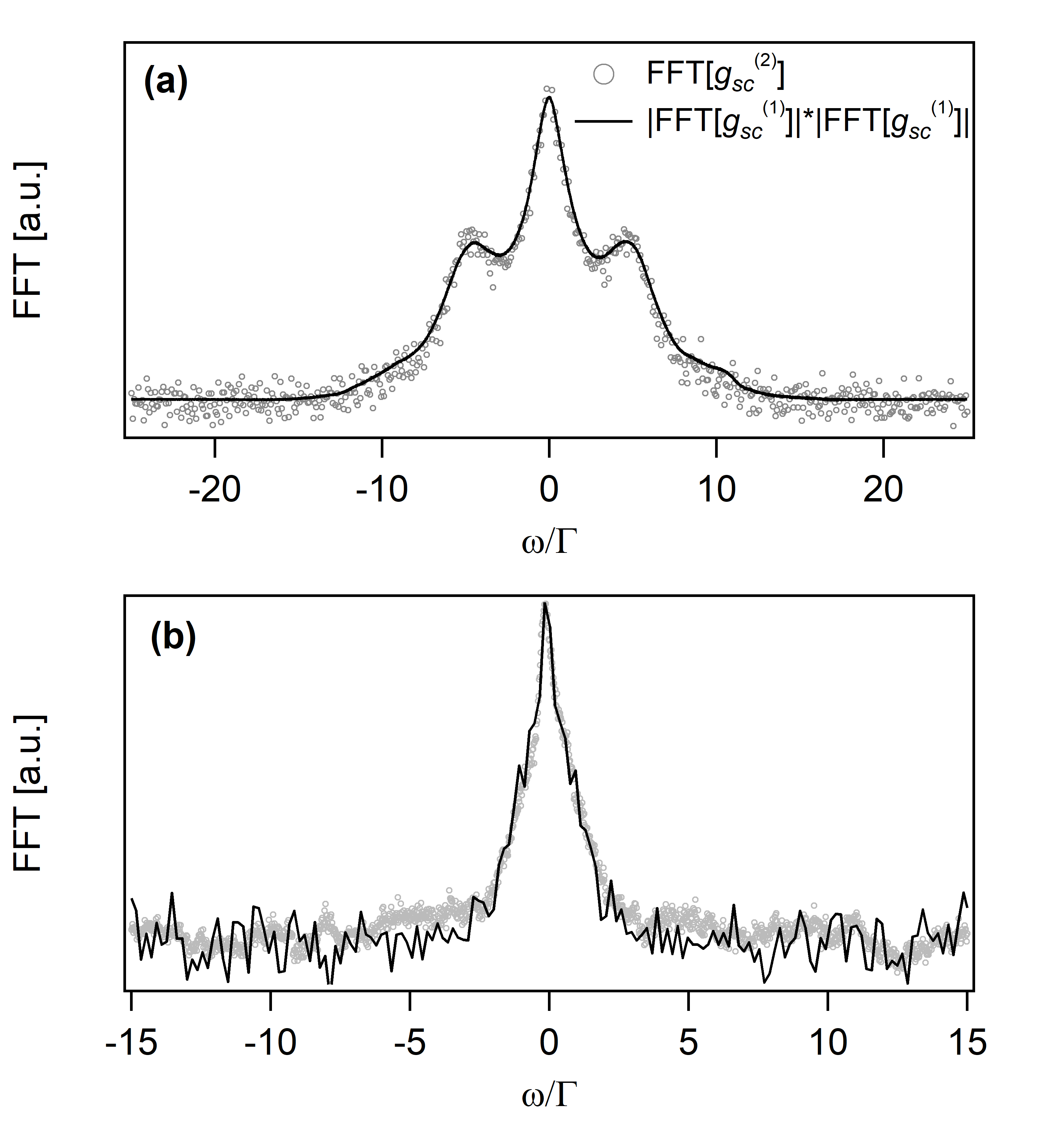}
	\caption{Siegert relation for light scattered by quantum scatterers (a) in the high saturation limit ($s=60$) and (b) in the intermediate regime ($s=1$). Open grey circles: Fourier transform of the intensity correlation $\tilde g_\mathrm{sc}{}^{(2)}$; plain black curve: self-convolution of the Fourier transform of the electric field correlation $\gomegasc * \tilde g_\mathrm{sc}{}^{(1)\star}(\omega)$.}
	\label{fig:Siegert_quantum}
\end{figure}

The two previous extracted quantities are plotted in Fig.\,\ref{fig:Siegert_quantum}a for $s=60$, thus with scattered light mainly composed of inelastic scattering. The open circles corresponds to $\tilde g_\mathrm{sc}{}^{(2)}(\omega)$, while the plain curve corresponds to $\gomegasc * \tilde g_\mathrm{sc}{}^{(1)\star}(\omega)$. One can see a very good overlap. We also performed the same measurement for $s=1$, for which we have equal contribution from elastic and inelastic scattering in terms of power. Fig.\,\ref{fig:Siegert_quantum}b presents the corresponding data, showing again a good overlap.

\begin{figure}[t]
	\centering
	\includegraphics[width=\columnwidth]{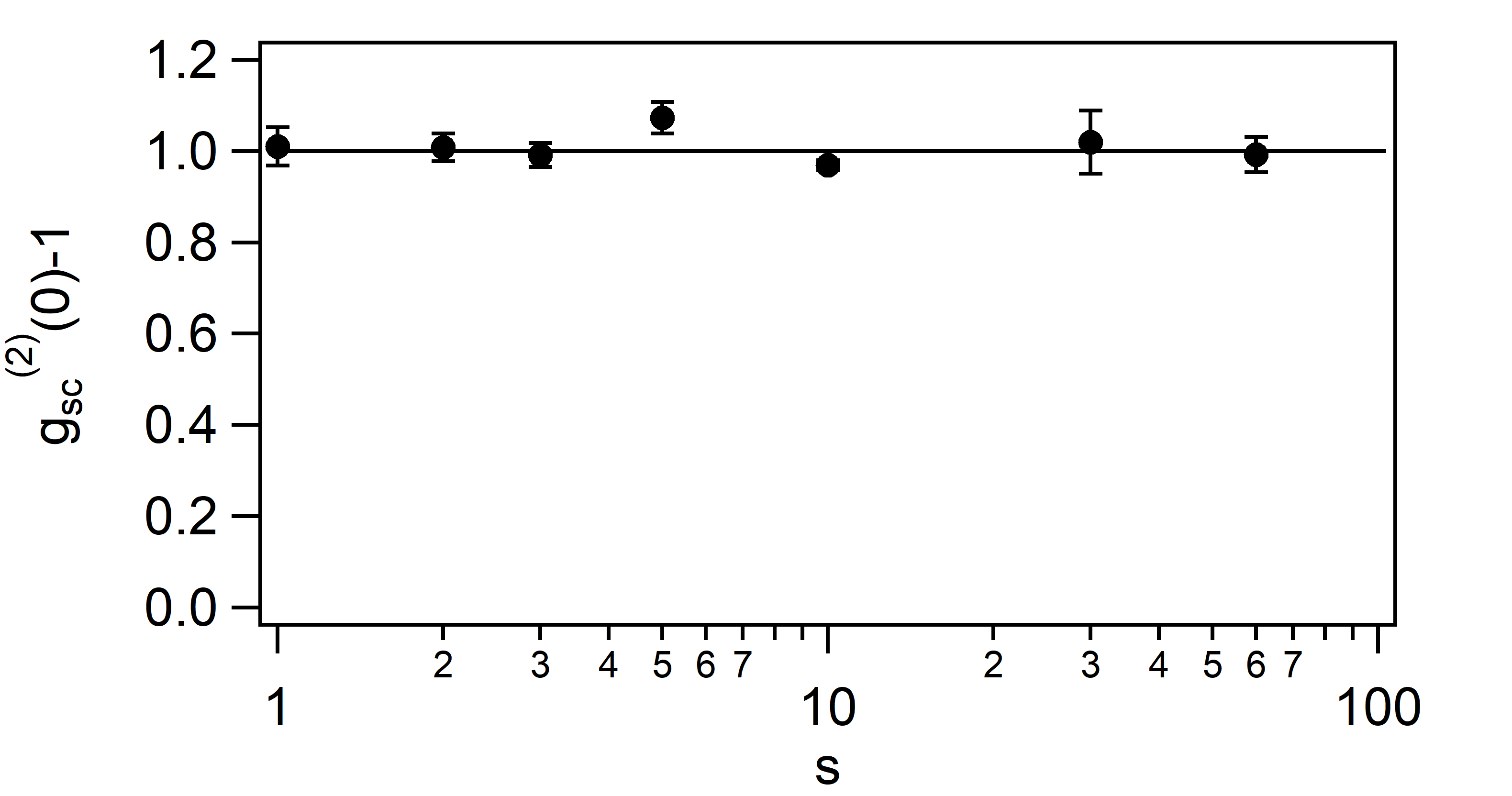}
	\caption{Contrast of the bunching peak for saturation parameters larger than one. The black line set at $g_\mathrm{sc}{}^{(2)}(0)-1 = 1$ corresponds to the expected value if the Siegert relation is valid.}
	\label{fig:g2_0_quantum}
\end{figure}

A good overlap between $\tilde g_\mathrm{sc}{}^{(2)}(\omega)$ and $\gomegasc * \tilde g_\mathrm{sc}{}^{(1)\star}(\omega)$ indicates that $\lvert g_\mathrm{sc}{}^{(1)}(\tau)\lvert^2$ and $g_\mathrm{sc}{}^{(2)}(\tau)$ have the same temporal shape, as expected from Eq.\,(\ref{eq:Siegert_simple}). However, to fully validate the Siegert relation, we also have to check the quantitative overlap, meaning the contrast of the bunching peak $g_\mathrm{sc}{}^{(2)}(0)-1$ should be equal to 1, the value of $g_\mathrm{sc}{}^{(1)}(0)$ being equal to 1 by definition. We thus take different measurements of $g_\mathrm{sc}{}^{(2)}(\tau)$ of the scattered light without the LO and for different values of $s$ between 1 and 60. The bunching peak contrast is plotted in Fig.\,\ref{fig:g2_0_quantum}. For most of the experimental values, the contrast, within the error bars, corresponds to the one expected if the Siegert relation is valid.

\section{Conclusion}\label{sec:conclusion}

In this paper, we have checked the validity of the Siegert relation in two different domains: in astronomy, for light coming from stars, and in cold-atom physics, when resonant laser light is scattered by a large number of quantum two-level systems. In astronomy, we have shown that this relation can be used to extract astrophysical information, such as the coherence time due to emission lines. Combined to spatial interferometry, this can be used to 
determine the angular diameter of the source. 
This is especially interesting to perform such measurements on emission lines to characterize the extended atmosphere of the star, which can give access to other fundamental parameters such as its distance \cite{Rivet:2020,Almeida2022}.
For cold atoms, we have shown that one can probe the quantum nature of the scatterers. 

To validate the Siegert relation, one needs to verify different assumptions. The first one is that the scatterers or emitters should be independent and their number should be large. The second assumption is that the emitted or scattered phase should be random and uncorrelated. While the process of phase randomization is obvious in stars with thermal radiation, it is a bit more complex for light scattered by quantum scatterers. The process that randomizes the phase actually depends on the scattering regime. On one hand, for low saturation parameter, scattering is mainly elastic and the phase randomization is due to temperature. The coherence time is thus linked to the atom velocities as well as the optical thickness which roughly corresponds to the number of scattering events that occurs inside the medium\,\cite{Eloy_2018}. On the other hand, when the saturation parameter is large, light is mainly inelastically scattered. The origin of the phase randomization is completely different since it comes from the finite lifetime of the two-level excited state. The coherence time is thus of the order of a few nanoseconds, much lower than the few hundreds of nanoseconds when due to temperature. We have shown in this paper that the Siegert relation is also valid in this configuration. As far as we know, this is the first time that the Siegert relation is checked with the simultaneous measurement of $\gtau$ and $\ggtau$ for large $s$ where the particles behave as quantum emitters.

The future of intensity correlations in astronomy can address two complementary goals. The first one is to increase the angular resolution in the visible, and more particularly for blue wavelength for which stellar amplitude interferometry\,\cite{Labeyrie:1975} is still hardly feasible. The second one is to look for deviation from the Siegert relation, with the ultimate goal of being able to have a direct signature of astrophysical lasing. On the quantum emitters side, one may search for deviation from the Siegert relation as a signature of correlations between the scatterers, like, e.g., in the specific experimental configurations of Ref.\,\cite{Prasad_2020}, where antibunching was observed with many-atoms trapped in the vicinity of a nanofiber. One could also look for signatures of random lasing in such samples, based on high optical thickness to induce multiple scattering and radiation trapping \cite{Labeyrie:2003}, and gain mechanisms such as Mollow gain\,\cite{Wu_1977, Guerin:2008, Froufe:2009}, Raman gain\,\cite{Tabosa_1991, Grison_1991, Hilico_1992, Guerin:2008, Guerin:2009, Vrijsen:2011, Baudouin:2013} or parametric gain with four-wave mixing\,\cite{Guerin:2008, Guerin_2010, Schilke:2012}.

\bmhead{Acknowledgments}

We acknowledge funding from the French National Research Agency (projects QuaCor, ANR19-CE47-0014-01 and I2C, ANR-20-CE31-0003). M.~A.~F.~B., A.~C. and R.~B. benefited from Grants from S\~ao Paulo Research Foundation (FAPESP, Grants Nos. 2018/15554-5, 2019/13143-0, 2017/09390-7 and 2021/02673-9) and from the National Council for Scientific and Technological Development (CNPq, Grant No.\,313886/2020-2). R.\,K., W.\,G., R.\,B. and M.\,H. received support from the project STIC-AmSud (Ph879-17/CAPES 88887.521971/2020-00). We also acknowledge the financial support of the UCA-JEDI project ANR-15-IDEX-01, the Doeblin Federation, the OPTIMAL platform, the Région PACA (project I2C). R.K. received support from the European project ANDLICA, ERC
Advanced Grant Agreement No. 832219.

\section*{Author contributions}

All authors contributed equally to the paper.

\section*{Data Availability}

The data underlying this article will be shared on reasonable request to the corresponding author.

\bibliography{Biblio,HBT_paper_biblio,Biblio_Siegert_Mollow}


\begin{thebibliography}{82}
\ifx \bisbn   \undefined \def \bisbn  #1{ISBN #1}\fi
\ifx \binits  \undefined \def \binits#1{#1}\fi
\ifx \bauthor  \undefined \def \bauthor#1{#1}\fi
\ifx \batitle  \undefined \def \batitle#1{#1}\fi
\ifx \bjtitle  \undefined \def \bjtitle#1{#1}\fi
\ifx \bvolume  \undefined \def \bvolume#1{\textbf{#1}}\fi
\ifx \byear  \undefined \def \byear#1{#1}\fi
\ifx \bissue  \undefined \def \bissue#1{#1}\fi
\ifx \bfpage  \undefined \def \bfpage#1{#1}\fi
\ifx \blpage  \undefined \def \blpage #1{#1}\fi
\ifx \burl  \undefined \def \burl#1{\textsf{#1}}\fi
\ifx \doiurl  \undefined \def \doiurl#1{\url{https://doi.org/#1}}\fi
\ifx \betal  \undefined \def \betal{\textit{et al.}}\fi
\ifx \binstitute  \undefined \def \binstitute#1{#1}\fi
\ifx \binstitutionaled  \undefined \def \binstitutionaled#1{#1}\fi
\ifx \bctitle  \undefined \def \bctitle#1{#1}\fi
\ifx \beditor  \undefined \def \beditor#1{#1}\fi
\ifx \bpublisher  \undefined \def \bpublisher#1{#1}\fi
\ifx \bbtitle  \undefined \def \bbtitle#1{#1}\fi
\ifx \bedition  \undefined \def \bedition#1{#1}\fi
\ifx \bseriesno  \undefined \def \bseriesno#1{#1}\fi
\ifx \blocation  \undefined \def \blocation#1{#1}\fi
\ifx \bsertitle  \undefined \def \bsertitle#1{#1}\fi
\ifx \bsnm \undefined \def \bsnm#1{#1}\fi
\ifx \bsuffix \undefined \def \bsuffix#1{#1}\fi
\ifx \bparticle \undefined \def \bparticle#1{#1}\fi
\ifx \barticle \undefined \def \barticle#1{#1}\fi
\bibcommenthead
\ifx \bconfdate \undefined \def \bconfdate #1{#1}\fi
\ifx \botherref \undefined \def \botherref #1{#1}\fi
\ifx \url \undefined \def \url#1{\textsf{#1}}\fi
\ifx \bchapter \undefined \def \bchapter#1{#1}\fi
\ifx \bbook \undefined \def \bbook#1{#1}\fi
\ifx \bcomment \undefined \def \bcomment#1{#1}\fi
\ifx \oauthor \undefined \def \oauthor#1{#1}\fi
\ifx \citeauthoryear \undefined \def \citeauthoryear#1{#1}\fi
\ifx \endbibitem  \undefined \def \endbibitem {}\fi
\ifx \bconflocation  \undefined \def \bconflocation#1{#1}\fi
\ifx \arxivurl  \undefined \def \arxivurl#1{\textsf{#1}}\fi
\csname PreBibitemsHook\endcsname

\bibitem{Wiener1930}
\begin{barticle}
\bauthor{\bsnm{Wiener}, \binits{N.}}:
\batitle{Generalized harmonic analysis}.
\bjtitle{Acta mathematica}
\bvolume{55},
\bfpage{117}--\blpage{258}
(\byear{1930})
\end{barticle}
\endbibitem

\bibitem{Khintchine1934}
\begin{barticle}
\bauthor{\bsnm{Khintchine}, \binits{A.}}:
\batitle{Korrelationstheorie der station{\"a}ren stochastischen prozesse}.
\bjtitle{Mathematische Annalen}
\bvolume{109}(\bissue{1}),
\bfpage{604}--\blpage{615}
(\byear{1934})
\end{barticle}
\endbibitem

\bibitem{Glauber:1963b}
\begin{barticle}
\bauthor{\bsnm{Glauber}, \binits{R.J.}}:
\batitle{The quantum theory of optical coherence}.
\bjtitle{Phys. Rev.}
\bvolume{130},
\bfpage{2529}
(\byear{1963})
\end{barticle}
\endbibitem

\bibitem{Siegert:1943}
\begin{botherref}
\oauthor{\bsnm{Siegert}, \binits{A.J.F.}}:
On the fluctuations in signals returned by many independently moving
  scatterers.
Report: Radiation laboratory,
Massachusetts Insitute of Technology
(1943)
\end{botherref}
\endbibitem

\bibitem{Maret_1987}
\begin{barticle}
\bauthor{\bsnm{Maret}, \binits{G.}},
\bauthor{\bsnm{Wolf}, \binits{P.E.}}:
\batitle{Multiple light scattering from disordered media. {T}he effect of
  brownian motion of scatterers}.
\bjtitle{Zeitschrift f{\"u}r Physik B Condensed Matter}
\bvolume{65}(\bissue{4}),
\bfpage{409}--\blpage{413}
(\byear{1987}).
\doiurl{10.1007/BF01303762}
\end{barticle}
\endbibitem

\bibitem{Pine_1988}
\begin{barticle}
\bauthor{\bsnm{Pine}, \binits{D.J.}},
\bauthor{\bsnm{Weitz}, \binits{D.A.}},
\bauthor{\bsnm{Chaikin}, \binits{P.M.}},
\bauthor{\bsnm{Herbolzheimer}, \binits{E.}}:
\batitle{Diffusing wave spectroscopy}.
\bjtitle{Phys. Rev. Lett.}
\bvolume{60},
\bfpage{1134}--\blpage{1137}
(\byear{1988}).
\doiurl{10.1103/PhysRevLett.60.1134}
\end{barticle}
\endbibitem

\bibitem{Loudon:book}
\begin{bbook}
\bauthor{\bsnm{Loudon}, \binits{R.}}:
\bbtitle{The Quantum Theory of Light}.
\bpublisher{Oxford Science Publications},
\blocation{New York}
(\byear{1973})
\end{bbook}
\endbibitem

\bibitem{Chu2012}
\begin{bbook}
\bauthor{\bsnm{Chu}, \binits{B.}}:
\bbtitle{Laser Light Scattering: Basic Principles and Practice}.
\bpublisher{Academic Press},
\blocation{Cambridge, Massachusetts}
(\byear{2012}).
\burl{https://www.elsevier.com/books/laser-light-scattering/chu/978-0-12-174551-6}
\end{bbook}
\endbibitem

\bibitem{Teich_1988}
\begin{botherref}
\oauthor{\bsnm{Teich}, \binits{M.C.}},
\oauthor{\bsnm{Saleh}, \binits{B.E.A.}}:
I photon bunching and antibunching.
Progress in Optics,
vol. 26,
pp. 1--104.
Elsevier
(1988).
\doiurl{10.1016/S0079-6638(08)70174-4}.
\url{http://www.sciencedirect.com/science/article/pii/S0079663808701744}
\end{botherref}
\endbibitem

\bibitem{akkermans_montambaux_2007}
\begin{bbook}
\bauthor{\bsnm{Akkermans}, \binits{E.}},
\bauthor{\bsnm{Montambaux}, \binits{G.}}:
\bbtitle{Mesoscopic Physics of Electrons and Photons}.
\bpublisher{Cambridge University Press},
\blocation{Cambridge}
(\byear{2007}).
\doiurl{10.1017/CBO9780511618833}
\end{bbook}
\endbibitem

\bibitem{HBT:1952}
\begin{barticle}
\bauthor{\bsnm{{Hanbury~Brown}}, \binits{R.}},
\bauthor{\bsnm{{Jennison}}, \binits{R.C.}},
\bauthor{\bsnm{{Das~Gupta}}, \binits{M.K.}}:
\batitle{{Apparent angular sizes of discrete radio sources}}.
\bjtitle{Nature}
\bvolume{170},
\bfpage{1061}--\blpage{1063}
(\byear{1952})
\end{barticle}
\endbibitem

\bibitem{HBT:1954}
\begin{barticle}
\bauthor{\bsnm{{Hanbury~Brown}}, \binits{R.}},
\bauthor{\bsnm{{Twiss}}, \binits{R.G.}}:
\batitle{{A New type of interferometer for use in radio astronomy}}.
\bjtitle{Phil. Mag.}
\bvolume{45},
\bfpage{663}--\blpage{682}
(\byear{1954}).
\doiurl{10.1080/14786440708520475}
\end{barticle}
\endbibitem

\bibitem{HBT:1956a}
\begin{barticle}
\bauthor{\bsnm{{Hanbury~Brown}}, \binits{R.}},
\bauthor{\bsnm{Twiss}, \binits{R.Q.}}:
\batitle{Correlation between photons in two coherent beams of light}.
\bjtitle{Nature}
\bvolume{177},
\bfpage{27}--\blpage{29}
(\byear{1956})
\end{barticle}
\endbibitem

\bibitem{HB:Boffin}
\begin{bbook}
\bauthor{\bsnm{{Hanbury~Brown}}, \binits{R.}}:
\bbtitle{Boffin: A Personal Story of the Early Days of Radar, Radio Astronomy
  and Quantum Optics}.
\bpublisher{IOP Publishing},
\blocation{BristolTeich}
(\byear{1991})
\end{bbook}
\endbibitem

\bibitem{Dirac:1930}
\begin{bbook}
\bauthor{\bsnm{Dirac}, \binits{P.A.M.}}:
\bbtitle{The Principles of Quantum Mechanics}.
\bpublisher{Oxford University Press},
\blocation{Oxford}
(\byear{1930})
\end{bbook}
\endbibitem

\bibitem{HBT:1956b}
\begin{barticle}
\bauthor{\bsnm{{Hanbury~Brown}}, \binits{R.}},
\bauthor{\bsnm{Twiss}, \binits{R.Q.}}:
\batitle{A test of a new type of stellar interferometer on {Sirius}}.
\bjtitle{Nature}
\bvolume{178},
\bfpage{1046}--\blpage{1048}
(\byear{1956})
\end{barticle}
\endbibitem

\bibitem{Brannen:1956}
\begin{barticle}
\bauthor{\bsnm{Brannen}, \binits{E.}},
\bauthor{\bsnm{Ferguson}, \binits{H.I.S.}}:
\batitle{{The question of correlation between photons in coherent light rays}}.
\bjtitle{Nature}
\bvolume{178},
\bfpage{481}--\blpage{482}
(\byear{1956})
\end{barticle}
\endbibitem

\bibitem{HBT:1956c}
\begin{barticle}
\bauthor{\bsnm{{Hanbury~Brown}}, \binits{R.}},
\bauthor{\bsnm{{Twiss}}, \binits{R.Q.}}:
\batitle{{The question of correlation between photons in coherent light rays}}.
\bjtitle{Nature}
\bvolume{178},
\bfpage{1447}--\blpage{1448}
(\byear{1956})
\end{barticle}
\endbibitem

\bibitem{Purcell:1956}
\begin{barticle}
\bauthor{\bsnm{Purcell}, \binits{E.M.}}:
\batitle{{The question of correlation between photons in coherent light rays}}.
\bjtitle{Nature}
\bvolume{178},
\bfpage{1449}--\blpage{1450}
(\byear{1956})
\end{barticle}
\endbibitem

\bibitem{HBT:1957a}
\begin{barticle}
\bauthor{\bsnm{{Hanbury~Brown}}, \binits{R.}},
\bauthor{\bsnm{{Twiss}}, \binits{R.Q.}}:
\batitle{{Interferometry of the intensity fluctuations in light. I. Basic
  theory: The correlation between photons in coherent beams of radiation}}.
\bjtitle{Proc. R. Soc. A}
\bvolume{242},
\bfpage{300}--\blpage{324}
(\byear{1957})
\end{barticle}
\endbibitem

\bibitem{Kahn:1958}
\begin{barticle}
\bauthor{\bsnm{Kahn}, \binits{F.D.}}:
\batitle{On photon coincidences and hanbury brown's interferometer}.
\bjtitle{Opt. Acta}
\bvolume{5}(\bissue{3-4}),
\bfpage{93}--\blpage{100}
(\byear{1958}).
\doiurl{10.1080/713826258}
\end{barticle}
\endbibitem

\bibitem{HBT:1957}
\begin{barticle}
\bauthor{\bsnm{{Twiss}}, \binits{R.Q.}},
\bauthor{\bsnm{{Little}}, \binits{A.G.}},
\bauthor{\bsnm{{Hanbury~Brown}}, \binits{R.}}:
\batitle{{Correlation between photons, in coherent beams of light, detected by
  a coincidence counting technique}}.
\bjtitle{Nature}
\bvolume{180},
\bfpage{324}--\blpage{326}
(\byear{1957})
\end{barticle}
\endbibitem

\bibitem{Fano:1961}
\begin{barticle}
\bauthor{\bsnm{Fano}, \binits{U.}}:
\batitle{{Quantum theory of interference effects in the mixing of light from
  phase-independent sources}}.
\bjtitle{Am. J. Phys.}
\bvolume{29},
\bfpage{539}--\blpage{545}
(\byear{1961})
\end{barticle}
\endbibitem

\bibitem{Glauber:2006}
\begin{barticle}
\bauthor{\bsnm{Glauber}, \binits{R.J.}}:
\batitle{Nobel lecture: One hundred years of light quanta}.
\bjtitle{Rev. Mod. Phys.}
\bvolume{78},
\bfpage{1267}
(\byear{2006})
\end{barticle}
\endbibitem

\bibitem{Glauber:1963a}
\begin{barticle}
\bauthor{\bsnm{Glauber}, \binits{R.J.}}:
\batitle{Photon correlations}.
\bjtitle{Phys. Rev. Lett}
\bvolume{10},
\bfpage{84}
(\byear{1963})
\end{barticle}
\endbibitem

\bibitem{Glauber:1963c}
\begin{barticle}
\bauthor{\bsnm{Glauber}, \binits{R.J.}}:
\batitle{Coherent and incoherent states of the radiation field}.
\bjtitle{Phys. Rev.}
\bvolume{131},
\bfpage{2766}
(\byear{1963})
\end{barticle}
\endbibitem

\bibitem{Arecchi:1966}
\begin{barticle}
\bauthor{\bsnm{Arecchi}, \binits{F.T.}},
\bauthor{\bsnm{Gatti}, \binits{E.}},
\bauthor{\bsnm{Sona}, \binits{A.}}:
\batitle{Time distribution of photons from coherent and {Gaussian} sources}.
\bjtitle{Phys. Lett.}
\bvolume{20},
\bfpage{27}--\blpage{29}
(\byear{1966})
\end{barticle}
\endbibitem

\bibitem{Aspect_1981}
\begin{barticle}
\bauthor{\bsnm{Aspect}, \binits{A.}},
\bauthor{\bsnm{Grangier}, \binits{P.}},
\bauthor{\bsnm{Roger}, \binits{G.}}:
\batitle{Experimental tests of realistic local theories via bell's theorem}.
\bjtitle{Phys. Rev. Lett.}
\bvolume{47},
\bfpage{460}--\blpage{463}
(\byear{1981}).
\doiurl{10.1103/PhysRevLett.47.460}
\end{barticle}
\endbibitem

\bibitem{Aspect_1982_2}
\begin{barticle}
\bauthor{\bsnm{Aspect}, \binits{A.}},
\bauthor{\bsnm{Grangier}, \binits{P.}},
\bauthor{\bsnm{Roger}, \binits{G.}}:
\batitle{Experimental realization of einstein-podolsky-rosen-bohm
  gedankenexperiment: A new violation of bell's inequalities}.
\bjtitle{Phys. Rev. Lett.}
\bvolume{49},
\bfpage{91}--\blpage{94}
(\byear{1982}).
\doiurl{10.1103/PhysRevLett.49.91}
\end{barticle}
\endbibitem

\bibitem{Aspect_1982}
\begin{barticle}
\bauthor{\bsnm{Aspect}, \binits{A.}},
\bauthor{\bsnm{Dalibard}, \binits{J.}},
\bauthor{\bsnm{Roger}, \binits{G.}}:
\batitle{Experimental test of bell's inequalities using time- varying
  analyzers}.
\bjtitle{Phys. Rev. Lett.}
\bvolume{49},
\bfpage{1804}--\blpage{1807}
(\byear{1982}).
\doiurl{10.1103/PhysRevLett.49.1804}
\end{barticle}
\endbibitem

\bibitem{Grangier_1985}
\begin{barticle}
\bauthor{\bsnm{Grangier}, \binits{P.}},
\bauthor{\bsnm{Aspect}, \binits{A.}},
\bauthor{\bsnm{Vigue}, \binits{J.}}:
\batitle{Quantum interference effect for two atoms radiating a single photon}.
\bjtitle{Phys. Rev. Lett.}
\bvolume{54},
\bfpage{418}--\blpage{421}
(\byear{1985}).
\doiurl{10.1103/PhysRevLett.54.418}
\end{barticle}
\endbibitem

\bibitem{HBT1974}
\begin{barticle}
\bauthor{\bsnm{Hanbury~Brown}, \binits{R.}},
\bauthor{\bsnm{Davis}, \binits{J.}},
\bauthor{\bsnm{Allen}, \binits{L.R.}}:
\batitle{The angular diameters of $32$~stars}.
\bjtitle{MNRAS}
\bvolume{167},
\bfpage{121}--\blpage{136}
(\byear{1974})
\end{barticle}
\endbibitem

\bibitem{Labeyrie:1975}
\begin{barticle}
\bauthor{\bsnm{Labeyrie}, \binits{A.}}:
\batitle{Interference fringes obtained on {Vega} with two optical telescopes}.
\bjtitle{ApJ.}
\bvolume{196},
\bfpage{71}--\blpage{75}
(\byear{1975})
\end{barticle}
\endbibitem

\bibitem{LeBohec:2006}
\begin{barticle}
\bauthor{\bsnm{LeBohec}, \binits{S.}},
\bauthor{\bsnm{Holder}, \binits{J.}}:
\batitle{Optical intensity interferometry with atomospheric {Cherenkov}
  telescope arrays}.
\bjtitle{ApJ}
\bvolume{649},
\bfpage{399}--\blpage{405}
(\byear{2006})
\end{barticle}
\endbibitem

\bibitem{Acciari:2020}
\begin{barticle}
\bauthor{\bsnm{Acciari}, \binits{V.A.}},
\bauthor{\bsnm{Bernardos}, \binits{M.I.}},
\bauthor{\bsnm{Colombo}, \binits{E.}},
\bauthor{\bsnm{Contreras}, \binits{J.L.}},
\bauthor{\bsnm{Cortina}, \binits{J.}},
\bauthor{\bsnm{De Angelis}, \binits{A.}},
\bauthor{\bsnm{Delgado}, \binits{C.}},
\bauthor{\bsnm{Díaz}, \binits{C.}},
\bauthor{\bsnm{Fink}, \binits{D.}},
\bauthor{\bsnm{Mariotti}, \binits{M.}},
\bauthor{\bsnm{Mangano}, \binits{S.}},
\bauthor{\bsnm{Mirzoyan}, \binits{R.}},
\bauthor{\bsnm{Polo}, \binits{M.}},
\bauthor{\bsnm{Schweizer}, \binits{T.}},
\bauthor{\bsnm{Will}, \binits{M.}}:
\batitle{{Optical intensity interferometry observations using the MAGIC Imaging
  Atmospheric Cherenkov Telescopes}}.
\bjtitle{MNRAS}
\bvolume{491}(\bissue{2}),
\bfpage{1540}--\blpage{1547}
(\byear{2020})
{\href{https://arxiv.org/abs/https://academic.oup.com/mnras/article-pdf/491/2/1540/31144895/stz3171.pdf}{{https://academic.oup.com/mnras/article-pdf/491/2/1540/31144895/stz3171.pdf}}}.
\doiurl{10.1093/mnras/stz3171}
\end{barticle}
\endbibitem

\bibitem{Abeysekara:2020}
\begin{barticle}
\bauthor{\bsnm{Abeysekara}, \binits{A.U.}},
\bauthor{\bsnm{Benbow}, \binits{W.}},
\bauthor{\bsnm{Brill}, \binits{A.}},
\bauthor{\bsnm{Buckley}, \binits{J.H.}},
\bauthor{\bsnm{Christiansen}, \binits{J.L.}},
\bauthor{\bsnm{Chromey}, \binits{A.J.}},
\bauthor{\bsnm{Daniel}, \binits{M.K.}},
\bauthor{\bsnm{Davis}, \binits{J.}},
\bauthor{\bsnm{Falcone}, \binits{A.}},
\bauthor{\bsnm{Feng}, \binits{Q.}},
\bauthor{\bsnm{Finley}, \binits{J.P.}},
\bauthor{\bsnm{Fortson}, \binits{L.}},
\bauthor{\bsnm{Furniss}, \binits{A.}},
\bauthor{\bsnm{Gent}, \binits{A.}},
\bauthor{\bsnm{Giuri}, \binits{C.}},
\bauthor{\bsnm{Gueta}, \binits{O.}},
\bauthor{\bsnm{Hanna}, \binits{D.}},
\bauthor{\bsnm{Hassan}, \binits{T.}},
\bauthor{\bsnm{Hervet}, \binits{O.}},
\bauthor{\bsnm{Holder}, \binits{J.}},
\bauthor{\bsnm{Hughes}, \binits{G.}},
\bauthor{\bsnm{Humensky}, \binits{T.B.}},
\bauthor{\bsnm{Kaaret}, \binits{P.}},
\bauthor{\bsnm{Kertzman}, \binits{M.}},
\bauthor{\bsnm{Kieda}, \binits{D.}},
\bauthor{\bsnm{Krennrich}, \binits{F.}},
\bauthor{\bsnm{Kumar}, \binits{S.}},
\bauthor{\bsnm{LeBohec}, \binits{T.}},
\bauthor{\bsnm{Lin}, \binits{T.T.Y.}},
\bauthor{\bsnm{Lundy}, \binits{M.}},
\bauthor{\bsnm{Maier}, \binits{G.}},
\bauthor{\bsnm{Matthews}, \binits{N.}},
\bauthor{\bsnm{Moriarty}, \binits{P.}},
\bauthor{\bsnm{Mukherjee}, \binits{R.}},
\bauthor{\bsnm{Nievas-Rosillo}, \binits{M.}},
\bauthor{\bsnm{O'Brien}, \binits{S.}},
\bauthor{\bsnm{Ong}, \binits{R.A.}},
\bauthor{\bsnm{Otte}, \binits{A.N.}},
\bauthor{\bsnm{Pfrang}, \binits{K.}},
\bauthor{\bsnm{Pohl}, \binits{M.}},
\bauthor{\bsnm{Prado}, \binits{R.R.}},
\bauthor{\bsnm{Pueschel}, \binits{E.}},
\bauthor{\bsnm{Quinn}, \binits{J.}},
\bauthor{\bsnm{Ragan}, \binits{K.}},
\bauthor{\bsnm{Reynolds}, \binits{P.T.}},
\bauthor{\bsnm{Ribeiro}, \binits{D.}},
\bauthor{\bsnm{Richards}, \binits{G.T.}},
\bauthor{\bsnm{Roache}, \binits{E.}},
\bauthor{\bsnm{Ryan}, \binits{J.L.}},
\bauthor{\bsnm{Santander}, \binits{M.}},
\bauthor{\bsnm{Sembroski}, \binits{G.H.}},
\bauthor{\bsnm{Wakely}, \binits{S.P.}},
\bauthor{\bsnm{Weinstein}, \binits{A.}},
\bauthor{\bsnm{Wilcox}, \binits{P.}},
\bauthor{\bsnm{Williams}, \binits{D.A.}},
\bauthor{\bsnm{Williamson}, \binits{T.J.}}:
\batitle{Demonstration of stellar intensity interferometry with the four
  {VERITAS} telescopes}.
\bjtitle{Nature Astronomy}
\bvolume{4}(\bissue{12}),
\bfpage{1164}--\blpage{1169}
(\byear{2020}).
\doiurl{10.1038/s41550-020-1143-y}
\end{barticle}
\endbibitem

\bibitem{Guerin:2018}
\begin{barticle}
\bauthor{\bsnm{Guerin}, \binits{W.}},
\bauthor{\bsnm{Rivet}, \binits{J.-P.}},
\bauthor{\bsnm{Fouch\'e}, \binits{M.}},
\bauthor{\bsnm{Labeyrie}, \binits{G.}},
\bauthor{\bsnm{Vernet}, \binits{D.}},
\bauthor{\bsnm{Vakili}, \binits{F.}},
\bauthor{\bsnm{Kaiser}, \binits{R.}}:
\batitle{Spatial intensity interferometry on three bright stars}.
\bjtitle{MNRAS}
\bvolume{480}(\bissue{4}),
\bfpage{245}--\blpage{250}
(\byear{2018}).
\doiurl{10.1093/mnras/stx2143}
\end{barticle}
\endbibitem

\bibitem{Rivet:2018}
\begin{barticle}
\bauthor{\bsnm{Rivet}, \binits{J.-P.}},
\bauthor{\bsnm{Vakili}, \binits{F.}},
\bauthor{\bsnm{Lai}, \binits{O.}},
\bauthor{\bsnm{Vernet}, \binits{D.}},
\bauthor{\bsnm{Fouch\'e}, \binits{M.}},
\bauthor{\bsnm{Guerin}, \binits{W.}},
\bauthor{\bsnm{Labeyrie}, \binits{G.}},
\bauthor{\bsnm{Kaiser}, \binits{R.}}:
\batitle{Optical long baseline intensity interferometry: prospects for stellar
  physics}.
\bjtitle{Exp. Astron.}
\bvolume{46},
\bfpage{531}--\blpage{542}
(\byear{2018}).
\doiurl{10.1007/s10686-018-9595-0}
\end{barticle}
\endbibitem

\bibitem{Lai:2018}
\begin{bchapter}
\bauthor{\bsnm{Lai}, \binits{O.}},
\bauthor{\bsnm{Guerin}, \binits{W.}},
\bauthor{\bsnm{Vakili}, \binits{F.}},
\bauthor{\bsnm{Kaiser}, \binits{R.}},
\bauthor{\bsnm{Rivet}, \binits{J.-P.}},
\bauthor{\bsnm{Fouch\'e}, \binits{M.}},
\bauthor{\bsnm{Labeyrie}, \binits{G.}},
\bauthor{\bsnm{Chab\'e}, \binits{J.}},
\bauthor{\bsnm{Courde}, \binits{C.}},
\bauthor{\bsnm{Samain}, \binits{E.}},
\bauthor{\bsnm{Vernet}, \binits{D.}}:
\bctitle{Intensity interferometry revival on the {C\^ote d'Azur}}.
In: \bbtitle{Proc. SPIE}.
\bsertitle{Optical and Infrared Interferometry and Imaging VI},
vol. \bseriesno{10701},
p. \bfpage{1070121}
(\byear{2018})
\end{bchapter}
\endbibitem

\bibitem{Matthews:2022}
\begin{bchapter}
\bauthor{\bsnm{Matthews}, \binits{N.}},
\bauthor{\bsnm{Rivet}, \binits{J.-P.}},
\bauthor{\bsnm{Hugbart}, \binits{M.}},
\bauthor{\bsnm{Labeyrie}, \binits{G.}},
\bauthor{\bsnm{Kaiser}, \binits{R.}},
\bauthor{\bsnm{Lai}, \binits{O.}},
\bauthor{\bsnm{Vakili}, \binits{F.}},
\bauthor{\bsnm{Vernet}, \binits{D.}},
\bauthor{\bsnm{Chab\'e}, \binits{J.}},
\bauthor{\bsnm{Courde}, \binits{C.}},
\bauthor{\bsnm{Schuler}, \binits{N.}},
\bauthor{\bsnm{Bourget}, \binits{P.}},
\bauthor{\bsnm{Guerin}, \binits{W.}}:
\bctitle{Intensity interferometry at calern and beyond: progress report}.
In: \bbtitle{Proc. SPIE}.
\bsertitle{Optical and Infrared Interferometry and Imaging VIII},
p.
(\byear{2022})
\end{bchapter}
\endbibitem

\bibitem{Goodman:book}
\begin{bbook}
\bauthor{\bsnm{Goodman}, \binits{J.W.}}:
\bbtitle{Speckle Phenomena in Optics: Theory and Applications},
\bedition{3rd} edn.
\bpublisher{Roberts \& Company Publishers},
\blocation{Greenwood Village}
(\byear{2009})
\end{bbook}
\endbibitem

\bibitem{Almeida2022}
\begin{barticle}
\bauthor{\bparticle{de} \bsnm{Almeida}, \binits{E.S.G.}},
\bauthor{\bsnm{Hugbart}, \binits{M.}},
\bauthor{\bparticle{de} \bsnm{Souza}, \binits{A.D.}},
\bauthor{\bsnm{Rivet}, \binits{J.-P.}},
\bauthor{\bsnm{Vakili}, \binits{F.}},
\bauthor{\bsnm{Siciak}, \binits{A.}},
\bauthor{\bsnm{Labeyrie}, \binits{G.}},
\bauthor{\bsnm{Garde}, \binits{O.}},
\bauthor{\bsnm{Matthews}, \binits{N.}},
\bauthor{\bsnm{Lai}, \binits{O.}},
\bauthor{\bsnm{Vernet}, \binits{D.}},
\bauthor{\bsnm{Kaiser}, \binits{R.}},
\bauthor{\bsnm{Guerin}, \binits{W.}}:
\batitle{{Combined spectroscopy and intensity interferometry to determine the
  distances of the blue supergiants P Cygni and Rigel}}.
\bjtitle{Monthly Notices of the Royal Astronomical Society}
(\byear{2022})
{\href{https://arxiv.org/abs/https://academic.oup.com/mnras/advance-article-pdf/doi/10.1093/mnras/stac1617/44082525/stac1617.pdf}{{https://academic.oup.com/mnras/advance-article-pdf/doi/10.1093/mnras/stac1617/44082525/stac1617.pdf}}}.
\doiurl{10.1093/mnras/stac1617}.
\bcomment{stac1617}
\end{barticle}
\endbibitem

\bibitem{Mandel1965}
\begin{barticle}
\bauthor{\bsnm{Mandel}, \binits{L.}},
\bauthor{\bsnm{Wolf}, \binits{E.}}:
\batitle{Coherence properties of optical fields}.
\bjtitle{Rev. Mod. Phys.}
\bvolume{37},
\bfpage{231}--\blpage{287}
(\byear{1965}).
\doiurl{10.1103/RevModPhys.37.231}
\end{barticle}
\endbibitem

\bibitem{Guerin:2017}
\begin{barticle}
\bauthor{\bsnm{Guerin}, \binits{W.}},
\bauthor{\bsnm{Dussaux}, \binits{A.}},
\bauthor{\bsnm{Fouch\'e}, \binits{M.}},
\bauthor{\bsnm{Labeyrie}, \binits{G.}},
\bauthor{\bsnm{Rivet}, \binits{J.-P.}},
\bauthor{\bsnm{Vernet}, \binits{D.}},
\bauthor{\bsnm{Vakili}, \binits{F.}},
\bauthor{\bsnm{Kaiser}, \binits{R.}}:
\batitle{Temporal intensity interferometry: photon bunching in three bright
  stars}.
\bjtitle{MNRAS}
\bvolume{472},
\bfpage{4126}--\blpage{4132}
(\byear{2017}).
\doiurl{10.1093/mnras/stx2143}
\end{barticle}
\endbibitem

\bibitem{AAVSO}
\begin{botherref}
\oauthor{\bsnm{AAVSO}}
AAVSO spectral database, Public data available at:
  \url{https://app.aavso.org/avspec/obs/7783}.
(2020)
\end{botherref}
\endbibitem

\bibitem{Rivet:2020}
\begin{barticle}
\bauthor{\bsnm{Rivet}, \binits{J.-P.}},
\bauthor{\bsnm{Siciak}, \binits{A.}},
\bauthor{\bsnm{{de~Almeida}}, \binits{E.S.G.}},
\bauthor{\bsnm{Vakili}, \binits{F.}},
\bauthor{\bsnm{{Domiciano de Souza}}, \binits{A.}},
\bauthor{\bsnm{Fouch\'e}, \binits{M.}},
\bauthor{\bsnm{Lai}, \binits{O.}},
\bauthor{\bsnm{Vernet}, \binits{D.}},
\bauthor{\bsnm{Kaiser}, \binits{R.}},
\bauthor{\bsnm{Guerin}, \binits{W.}}:
\batitle{Intensity interferometry of {P\,Cygni} in the {H}$\alpha$ emission
  line: towards distance calibration of {LBV} supergiant stars}.
\bjtitle{MNRAS}
\bvolume{494},
\bfpage{218}--\blpage{2020}
(\byear{2020}).
\doiurl{10.1093/mnras/staa588}
\end{barticle}
\endbibitem

\bibitem{Johansson:2007}
\begin{barticle}
\bauthor{\bsnm{Johansson}, \binits{S.}},
\bauthor{\bsnm{Letokhov}, \binits{V.S.}}:
\batitle{Astrophysical lasers and nonlinear optical effects in space}.
\bjtitle{New. Astron. Rev.}
\bvolume{51},
\bfpage{443}--\blpage{523}
(\byear{2007})
\end{barticle}
\endbibitem

\bibitem{Reid:1981}
\begin{barticle}
\bauthor{\bsnm{Reid}, \binits{M.J.}},
\bauthor{\bsnm{Moran}, \binits{J.M.}}:
\batitle{Masers}.
\bjtitle{Ann. Rev. Astron. Astrophys.}
\bvolume{19},
\bfpage{231}--\blpage{276}
(\byear{1981})
\end{barticle}
\endbibitem

\bibitem{Elitzur:1982}
\begin{barticle}
\bauthor{\bsnm{Elitzur}, \binits{M.}}:
\batitle{Physical characteristics of astronomical masers}.
\bjtitle{Rev. Mod. Phys.}
\bvolume{54},
\bfpage{1225}--\blpage{1260}
(\byear{1982})
\end{barticle}
\endbibitem

\bibitem{Johnson:1976}
\begin{barticle}
\bauthor{\bsnm{Johnson}, \binits{M.A.}},
\bauthor{\bsnm{Betz}, \binits{M.A.}},
\bauthor{\bsnm{McLaren}, \binits{R.A.}},
\bauthor{\bsnm{Sutton}, \binits{E.C.}},
\bauthor{\bsnm{Townes}, \binits{C.H.}}:
\batitle{Nonthermal 10 micron {CO$_2$} emission lines in the atmospheres of
  {Mars} and {Venus}}.
\bjtitle{ApJ}
\bvolume{208},
\bfpage{145}--\blpage{148}
(\byear{1976})
\end{barticle}
\endbibitem

\bibitem{Mumma:1981}
\begin{barticle}
\bauthor{\bsnm{Mumma}, \binits{M.J.}},
\bauthor{\bsnm{Buhl}, \binits{D.}},
\bauthor{\bsnm{Chin}, \binits{G.}},
\bauthor{\bsnm{Deming}, \binits{D.}},
\bauthor{\bsnm{Espenak}, \binits{F.}},
\bauthor{\bsnm{Kostiuk}, \binits{T.}},
\bauthor{\bsnm{Zipoy}, \binits{D.}}:
\batitle{Discovery of natural gain amplification in the 10-micrometer carbon
  dioxide laser bands on {Mars}: a natural laser}.
\bjtitle{Science}
\bvolume{212},
\bfpage{45}--\blpage{49}
(\byear{1981})
\end{barticle}
\endbibitem

\bibitem{Strelnitski:1996}
\begin{barticle}
\bauthor{\bsnm{Strelnitski}, \binits{V.S.}},
\bauthor{\bsnm{Haas}, \binits{M.R.}},
\bauthor{\bsnm{Smith}, \binits{H.A.}},
\bauthor{\bsnm{Erickson}, \binits{E.F.}},
\bauthor{\bsnm{Colgan}, \binits{S.W.J.}},
\bauthor{\bsnm{Hallenbach}, \binits{D.J.}}:
\batitle{Far-infrared hydrogen lasers in the peculiar star {MWC 349A}}.
\bjtitle{Science}
\bvolume{272},
\bfpage{1459}--\blpage{1461}
(\byear{1996})
\end{barticle}
\endbibitem

\bibitem{Johansson:2004}
\begin{barticle}
\bauthor{\bsnm{Johansson}, \binits{S.}},
\bauthor{\bsnm{Letokhov}, \binits{V.S.}}:
\batitle{Astrophysical lasers operating in optical {Fe II} lines in stellar
  ejecta of $\eta$ carinae}.
\bjtitle{A\&A}
\bvolume{428},
\bfpage{497}--\blpage{509}
(\byear{2004})
\end{barticle}
\endbibitem

\bibitem{Johansson:2005}
\begin{barticle}
\bauthor{\bsnm{Johansson}, \binits{S.}},
\bauthor{\bsnm{Letokhov}, \binits{V.S.}}:
\batitle{Astrophysical laser operating in the {O I 8446-{\AA}} line in the
  weigelt blobs of $\eta$ {Carinae}}.
\bjtitle{Mon. Not. Roy. Astron. Soc.}
\bvolume{364},
\bfpage{731}--\blpage{737}
(\byear{2005})
\end{barticle}
\endbibitem

\bibitem{Messenger:2010}
\begin{barticle}
\bauthor{\bsnm{Messenger}, \binits{S.J.}},
\bauthor{\bsnm{Strelnitski}, \binits{V.}}:
\batitle{On the 1.7 $\mu$m {Fe II} and other natural lasers}.
\bjtitle{Mon. Not. Roy. Astron. Soc.}
\bvolume{404},
\bfpage{1545}--\blpage{1550}
(\byear{2010})
\end{barticle}
\endbibitem

\bibitem{Lavrinovich:1975}
\begin{barticle}
\bauthor{\bsnm{Lavrinovich}, \binits{N.N.}},
\bauthor{\bsnm{Letokhov}, \binits{V.S.}}:
\batitle{The possibility of the laser effect in stellar atmospheres}.
\bjtitle{Sov. Phys.-JETP}
\bvolume{40},
\bfpage{800}
(\byear{1975})
\end{barticle}
\endbibitem

\bibitem{Wiersma:2008}
\begin{barticle}
\bauthor{\bsnm{Wiersma}, \binits{D.S.}}:
\batitle{The physics and applications of random lasers}.
\bjtitle{Nature Phys.}
\bvolume{4},
\bfpage{359}--\blpage{367}
(\byear{2008})
\end{barticle}
\endbibitem

\bibitem{Baudouin:2013}
\begin{barticle}
\bauthor{\bsnm{Baudouin}, \binits{Q.}},
\bauthor{\bsnm{Mercadier}, \binits{N.}},
\bauthor{\bsnm{Guarrera}, \binits{V.}},
\bauthor{\bsnm{Guerin}, \binits{W.}},
\bauthor{\bsnm{Kaiser}, \binits{R.}}:
\batitle{A cold-atom random laser}.
\bjtitle{Nature Phys.}
\bvolume{9},
\bfpage{357}--\blpage{360}
(\byear{2013})
\end{barticle}
\endbibitem

\bibitem{Cao_2001}
\begin{barticle}
\bauthor{\bsnm{Cao}, \binits{H.}},
\bauthor{\bsnm{Ling}, \binits{Y.}},
\bauthor{\bsnm{Xu}, \binits{J.Y.}},
\bauthor{\bsnm{Cao}, \binits{C.Q.}},
\bauthor{\bsnm{Kumar}, \binits{P.}}:
\batitle{Photon statistics of random lasers with resonant feedback}.
\bjtitle{Phys. Rev. Lett.}
\bvolume{86},
\bfpage{4524}--\blpage{4527}
(\byear{2001}).
\doiurl{10.1103/PhysRevLett.86.4524}
\end{barticle}
\endbibitem

\bibitem{Raposo_2022}
\begin{barticle}
\bauthor{\bsnm{Raposo}, \binits{E.P.}},
\bauthor{\bsnm{Gonz\'alez}, \binits{I.R.R.}},
\bauthor{\bsnm{Coronel}, \binits{E.D.}},
\bauthor{\bsnm{Mac\^edo}, \binits{A.M.S.}},
\bauthor{\bsnm{Menezes}, \binits{L.d.S.}},
\bauthor{\bsnm{Kashyap}, \binits{R.}},
\bauthor{\bsnm{Gomes}, \binits{A.S.L.}},
\bauthor{\bsnm{Kaiser}, \binits{R.}}:
\batitle{Intensity ${g}^{(2)}$ correlations in random fiber lasers: A
  random-matrix-theory approach}.
\bjtitle{Phys. Rev. A}
\bvolume{105},
\bfpage{031502}
(\byear{2022}).
\doiurl{10.1103/PhysRevA.105.L031502}
\end{barticle}
\endbibitem

\bibitem{Eloy_2018}
\begin{barticle}
\bauthor{\bsnm{Eloy}, \binits{A.}},
\bauthor{\bsnm{Yao}, \binits{Z.}},
\bauthor{\bsnm{Bachelard}, \binits{R.}},
\bauthor{\bsnm{Guerin}, \binits{W.}},
\bauthor{\bsnm{Fouch\'e}, \binits{M.}},
\bauthor{\bsnm{Kaiser}, \binits{R.}}:
\batitle{Diffusing-wave spectroscopy of cold atoms in ballistic motion}.
\bjtitle{Phys. Rev. A}
\bvolume{97},
\bfpage{013810}
(\byear{2018}).
\doiurl{10.1103/PhysRevA.97.013810}
\end{barticle}
\endbibitem

\bibitem{Kuusela_2017}
\begin{barticle}
\bauthor{\bsnm{Kuusela}, \binits{T.A.}}:
\batitle{Measurement of the second-order coherence of pseudothermal light}.
\bjtitle{American Journal of Physics}
\bvolume{85}(\bissue{4}),
\bfpage{289}--\blpage{294}
(\byear{2017}).
\doiurl{10.1119/1.4975212}
\end{barticle}
\endbibitem

\bibitem{Ferreira_2020}
\begin{barticle}
\bauthor{\bsnm{Ferreira}, \binits{D.}},
\bauthor{\bsnm{Bachelard}, \binits{R.}},
\bauthor{\bsnm{Guerin}, \binits{W.}},
\bauthor{\bsnm{Kaiser}, \binits{R.}},
\bauthor{\bsnm{Fouché}, \binits{M.}}:
\batitle{Connecting field and intensity correlations: The siegert relation and
  how to test it}.
\bjtitle{American Journal of Physics}
\bvolume{88}(\bissue{10}),
\bfpage{831}--\blpage{837}
(\byear{2020})
{\href{https://arxiv.org/abs/https://doi.org/10.1119/10.0001630}{{https://doi.org/10.1119/10.0001630}}}.
\doiurl{10.1119/10.0001630}
\end{barticle}
\endbibitem

\bibitem{Mollow_1969}
\begin{barticle}
\bauthor{\bsnm{Mollow}, \binits{B.R.}}:
\batitle{Power spectrum of light scattered by two-level systems}.
\bjtitle{Phys. Rev.}
\bvolume{188},
\bfpage{1969}--\blpage{1975}
(\byear{1969}).
\doiurl{10.1103/PhysRev.188.1969}
\end{barticle}
\endbibitem

\bibitem{Schrama_1992}
\begin{barticle}
\bauthor{\bsnm{Schrama}, \binits{C.A.}},
\bauthor{\bsnm{Nienhuis}, \binits{G.}},
\bauthor{\bsnm{Dijkerman}, \binits{H.A.}},
\bauthor{\bsnm{Steijsiger}, \binits{C.}},
\bauthor{\bsnm{Heideman}, \binits{H.G.M.}}:
\batitle{Intensity correlations between the components of the resonance
  fluorescence triplet}.
\bjtitle{Phys. Rev. A}
\bvolume{45},
\bfpage{8045}--\blpage{8055}
(\byear{1992}).
\doiurl{10.1103/PhysRevA.45.8045}
\end{barticle}
\endbibitem

\bibitem{Aspect_1980}
\begin{barticle}
\bauthor{\bsnm{Aspect}, \binits{A.}},
\bauthor{\bsnm{Roger}, \binits{G.}},
\bauthor{\bsnm{Reynaud}, \binits{S.}},
\bauthor{\bsnm{Dalibard}, \binits{J.}},
\bauthor{\bsnm{Cohen-Tannoudji}, \binits{C.}}:
\batitle{Time correlations between the two sidebands of the resonance
  fluorescence triplet}.
\bjtitle{Phys. Rev. Lett.}
\bvolume{45},
\bfpage{617}--\blpage{620}
(\byear{1980}).
\doiurl{10.1103/PhysRevLett.45.617}
\end{barticle}
\endbibitem

\bibitem{Grangier_1986}
\begin{barticle}
\bauthor{\bsnm{Grangier}, \binits{P.}},
\bauthor{\bsnm{Roger}, \binits{G.}},
\bauthor{\bsnm{Aspect}, \binits{A.}},
\bauthor{\bsnm{Heidmann}, \binits{A.}},
\bauthor{\bsnm{Reynaud}, \binits{S.}}:
\batitle{Observation of photon antibunching in phase-matched multiatom
  resonance fluorescence}.
\bjtitle{Phys. Rev. Lett.}
\bvolume{57},
\bfpage{687}--\blpage{690}
(\byear{1986}).
\doiurl{10.1103/PhysRevLett.57.687}
\end{barticle}
\endbibitem

\bibitem{steck2007quantum}
\begin{bbook}
\bauthor{\bsnm{Steck}, \binits{D.A.}}:
\bbtitle{Quantum and Atom Optics},
(\byear{2007}).
\burl{http://steck.us/teaching}
\end{bbook}
\endbibitem

\bibitem{Kashanian_2016}
\begin{barticle}
\bauthor{\bsnm{Kashanian}, \binits{S.V.}},
\bauthor{\bsnm{Eloy}, \binits{A.}},
\bauthor{\bsnm{Guerin}, \binits{W.}},
\bauthor{\bsnm{Lintz}, \binits{M.}},
\bauthor{\bsnm{Fouch\'e}, \binits{M.}},
\bauthor{\bsnm{Kaiser}, \binits{R.}}:
\batitle{Noise spectroscopy with large clouds of cold atoms}.
\bjtitle{Phys. Rev. A}
\bvolume{94},
\bfpage{043622}
(\byear{2016}).
\doiurl{10.1103/PhysRevA.94.043622}
\end{barticle}
\endbibitem

\bibitem{Ortiz_2019}
\begin{barticle}
\bauthor{\bsnm{Ortiz}, \binits{L.}},
\bauthor{\bsnm{Teixeira}, \binits{R.C.}},
\bauthor{\bsnm{Eloy}, \binits{A.}},
\bauthor{\bsnm{Ferreira}, \binits{D.}},
\bauthor{\bsnm{Kaiser}, \binits{R.}},
\bauthor{\bsnm{Bachelard}, \binits{R.}},
\bauthor{\bsnm{Fouch\'e}, \binits{M.}}:
\batitle{Mollow triplet in cold atoms}.
\bjtitle{New Journal of Physics}
\bvolume{21},
\bfpage{093019}
(\byear{2019})
\end{barticle}
\endbibitem

\bibitem{Prasad_2020}
\begin{barticle}
\bauthor{\bsnm{Prasad}, \binits{A.S.}},
\bauthor{\bsnm{Hinney}, \binits{J.}},
\bauthor{\bsnm{Mahmoodian}, \binits{S.}},
\bauthor{\bsnm{Hammerer}, \binits{K.}},
\bauthor{\bsnm{Rind}, \binits{S.}},
\bauthor{\bsnm{Schneeweiss}, \binits{P.}},
\bauthor{\bsnm{Sørensen}, \binits{A.S.}},
\bauthor{\bsnm{Volz}, \binits{J.}},
\bauthor{\bsnm{Rauschenbeutel}, \binits{A.}}:
\batitle{Correlating photons using the collective nonlinear response of atoms
  weakly coupled to an optical mode}.
\bjtitle{Nat. Photonics}
\bvolume{14},
\bfpage{719}--\blpage{722}
(\byear{2020})
\end{barticle}
\endbibitem

\bibitem{Labeyrie:2003}
\begin{barticle}
\bauthor{\bsnm{Labeyrie}, \binits{G.}},
\bauthor{\bsnm{Vaujour}, \binits{E.}},
\bauthor{\bsnm{M\"uller}, \binits{C.A.}},
\bauthor{\bsnm{Delande}, \binits{D.}},
\bauthor{\bsnm{Miniatura}, \binits{C.}},
\bauthor{\bsnm{Wilkowski}, \binits{D.}},
\bauthor{\bsnm{Kaiser}, \binits{R.}}:
\batitle{Slow diffusion of light in a cold atomic cloud}.
\bjtitle{Phys. Rev. Lett.}
\bvolume{91},
\bfpage{223904}
(\byear{2003})
\end{barticle}
\endbibitem

\bibitem{Wu_1977}
\begin{barticle}
\bauthor{\bsnm{Wu}, \binits{F.Y.}},
\bauthor{\bsnm{Ezekiel}, \binits{S.}},
\bauthor{\bsnm{Ducloy}, \binits{M.}},
\bauthor{\bsnm{Mollow}, \binits{B.R.}}:
\batitle{Observation of amplification in a strongly driven two-level atomic
  system at optical frequencies}.
\bjtitle{Phys. Rev. Lett.}
\bvolume{38},
\bfpage{1077}--\blpage{1080}
(\byear{1977}).
\doiurl{10.1103/PhysRevLett.38.1077}
\end{barticle}
\endbibitem

\bibitem{Guerin:2008}
\begin{barticle}
\bauthor{\bsnm{Guerin}, \binits{W.}},
\bauthor{\bsnm{Michaud}, \binits{F.}},
\bauthor{\bsnm{Kaiser}, \binits{R.}}:
\batitle{Mechanisms for lasing with cold atoms as the gain medium}.
\bjtitle{Phys. Rev. Lett.}
\bvolume{101},
\bfpage{093002}
(\byear{2008})
\end{barticle}
\endbibitem

\bibitem{Froufe:2009}
\begin{barticle}
\bauthor{\bsnm{Froufe-P\'erez}, \binits{L.S.}},
\bauthor{\bsnm{Guerin}, \binits{W.}},
\bauthor{\bsnm{Carminati}, \binits{R.}},
\bauthor{\bsnm{Kaiser}, \binits{R.}}:
\batitle{Threshold of a random laser with cold atoms}.
\bjtitle{Phys. Rev. Lett.}
\bvolume{102},
\bfpage{173903}
(\byear{2009})
\end{barticle}
\endbibitem

\bibitem{Tabosa_1991}
\begin{barticle}
\bauthor{\bsnm{Tabosa}, \binits{J.W.R.}},
\bauthor{\bsnm{Chen}, \binits{G.}},
\bauthor{\bsnm{Hu}, \binits{Z.}},
\bauthor{\bsnm{Lee}, \binits{R.B.}},
\bauthor{\bsnm{Kimble}, \binits{H.J.}}:
\batitle{Nonlinear spectroscopy of cold atoms in a spontaneous-force optical
  trap}.
\bjtitle{Phys. Rev. Lett.}
\bvolume{66},
\bfpage{3245}--\blpage{3248}
(\byear{1991}).
\doiurl{10.1103/PhysRevLett.66.3245}
\end{barticle}
\endbibitem

\bibitem{Grison_1991}
\begin{barticle}
\bauthor{\bsnm{Grison}, \binits{D.}},
\bauthor{\bsnm{Lounis}, \binits{B.}},
\bauthor{\bsnm{Salomon}, \binits{C.}},
\bauthor{\bsnm{Courtois}, \binits{J.Y.}},
\bauthor{\bsnm{Grynberg}, \binits{G.}}:
\batitle{Raman spectroscopy of cesium atoms in a laser trap}.
\bjtitle{Europhysics Letters ({EPL})}
\bvolume{15}(\bissue{2}),
\bfpage{149}--\blpage{154}
(\byear{1991}).
\doiurl{10.1209/0295-5075/15/2/007}
\end{barticle}
\endbibitem

\bibitem{Hilico_1992}
\begin{barticle}
\bauthor{\bsnm{Hilico}, \binits{L.}},
\bauthor{\bsnm{Fabre}, \binits{C.}},
\bauthor{\bsnm{Giacobino}, \binits{E.}}:
\batitle{Operation of a {\textquotedblleft}cold-atom laser{\textquotedblright}
  in a magneto-optical trap}.
\bjtitle{Europhysics Letters ({EPL})}
\bvolume{18}(\bissue{8}),
\bfpage{685}--\blpage{688}
(\byear{1992}).
\doiurl{10.1209/0295-5075/18/8/004}
\end{barticle}
\endbibitem

\bibitem{Guerin:2009}
\begin{barticle}
\bauthor{\bsnm{Guerin}, \binits{W.}},
\bauthor{\bsnm{Mercadier}, \binits{N.}},
\bauthor{\bsnm{Brivio}, \binits{D.}},
\bauthor{\bsnm{Kaiser}, \binits{R.}}:
\batitle{Threshold of a random laser based on {Raman} gain in cold atoms}.
\bjtitle{Opt. Express}
\bvolume{17},
\bfpage{11236}
(\byear{2009})
\end{barticle}
\endbibitem

\bibitem{Vrijsen:2011}
\begin{barticle}
\bauthor{\bsnm{Vrijsen}, \binits{G.}},
\bauthor{\bsnm{Hosten}, \binits{O.}},
\bauthor{\bsnm{Lee}, \binits{J.}},
\bauthor{\bsnm{Bernon}, \binits{S.}},
\bauthor{\bsnm{Kasevich}, \binits{M.A.}}:
\batitle{Raman lasing with a cold atom gain medium in a high-finesse optical
  cavity}.
\bjtitle{Phys. Rev. Lett.}
\bvolume{107},
\bfpage{063904}
(\byear{2011})
\end{barticle}
\endbibitem

\bibitem{Guerin_2010}
\begin{barticle}
\bauthor{\bsnm{Guerin}, \binits{W.}},
\bauthor{\bsnm{Mercadier}, \binits{N.}},
\bauthor{\bsnm{Michaud}, \binits{F.}},
\bauthor{\bsnm{Brivio}, \binits{D.}},
\bauthor{\bsnm{Froufe-P{\'{e}}rez}, \binits{L.S.}},
\bauthor{\bsnm{Carminati}, \binits{R.}},
\bauthor{\bsnm{Eremeev}, \binits{V.}},
\bauthor{\bsnm{Goetschy}, \binits{A.}},
\bauthor{\bsnm{Skipetrov}, \binits{S.E.}},
\bauthor{\bsnm{Kaiser}, \binits{R.}}:
\batitle{Towards a random laser with cold atoms}.
\bjtitle{Journal of Optics}
\bvolume{12}(\bissue{2}),
\bfpage{024002}
(\byear{2010}).
\doiurl{10.1088/2040-8978/12/2/024002}
\end{barticle}
\endbibitem

\bibitem{Schilke:2012}
\begin{barticle}
\bauthor{\bsnm{Schilke}, \binits{A.}},
\bauthor{\bsnm{Zimmermann}, \binits{C.}},
\bauthor{\bsnm{Courteille}, \binits{P.W.}},
\bauthor{\bsnm{Guerin}, \binits{W.}}:
\batitle{Optical parametric oscillation with distributed feedback in cold
  atoms}.
\bjtitle{Nature Photon.}
\bvolume{6},
\bfpage{101}
(\byear{2012})
\end{barticle}
\endbibitem

\end{thebibliography}

\end{document}